\documentclass[usenatbib]{mn2e}

\usepackage{psfig,epsfig,placeins,graphics,supertabular,wrapfig}
\usepackage{graphicx,subfigure}
\usepackage{marvosym,multirow,units}
\usepackage{graphics}
\usepackage{amssymb,amsmath}
\usepackage{euscript}
\usepackage{setspace,placeins}
\usepackage{afterpage}
\usepackage{times,rotating,url}
\usepackage{lscape}

\newcommand{\hatlas}{{\it H}-ATLAS}

\newcommand{\gtsim}{\mbox{{\raisebox{-0.4ex}{$\stackrel{>}{{\scriptstyle\sim}}$}
}}}
\newcommand{\ltsim}{\mbox{{\raisebox{-0.4ex}{$\stackrel{<}{{\scriptstyle\sim}}$}
}}}

\newcommand{\tgrey}{\ensuremath{T_{\mathrm{eff}}}}
\newcommand{\teff}{\ensuremath{T_{\mathrm{eff}}}}
\newcommand{\beff}{\ensuremath{\beta_{\mathrm{eff}}}}

\newcommand{\ldust}{\ensuremath{L_{\mathrm{dust}}}}

\title[Isothermal dust models of {\it H}-ATLAS galaxies]{Isothermal
  dust models of {\it Herschel}-ATLAS\thanks{{\it Herschel} is an ESA
    space observatory with science instruments provided by
    European-led Principal Investigator consortia and with important
    participation from NASA} galaxies}

\author[D.J.B.~Smith et al.]
       {D.J.B. Smith$^{1}$\thanks{E-mail:daniel.j.b.smith@gmail.com(DS)}, M.J. Hardcastle$^1$, M.J. Jarvis$^{2,3}$, S.J. Maddox$^4$, L. Dunne$^4$, \newauthor D.G. Bonfield$^1$, S. Eales$^5$, S. Serjeant$^6$, M.A.Thompson$^1$, M. Baes$^7$, D.L. Clements$^8$, \newauthor A. Cooray$^9$, G. De Zotti$^{10,11}$, J. Gonz\`alez-Nuevo$^{12}$, P. van der Werf$^{13}$, J. Virdee$^{2,14}$ \newauthor N. Bourne$^{15}$, A. Dariush$^{16}$, R. Hopwood$^{8,6}$, E. Ibar$^{17,18}$ \&\ E. Valiante$^{5}$       \vspace{0.05cm}\\
    $^{1}$Centre for Astrophysics Research, University of Hertfordshire, Hatfield, Herts, AL10 9AB, UK \\
    $^2$Department of Astrophysics, Denys Wilkinson Building, Keble Road, Oxford, OX1 3RH, UK \\
    $^{3}$Physics Department, University of the Western Cape, Private Bag X17, Bellville 7535, South Africa \\
    $^{4}$Dept of Physics and Astronomy, University of Canterbury, Private Bag 4800, Christchurch 8140, New Zealand \\
    $^{5}$School of Physics and Astronomy, Cardiff University, Queen's Buildings, The Parade, Cardiff CF24 3AA, UK \\
    $^{6}$Department of Physical Sciences, The Open University, Milton Keynes, MK7 6AA, UK \\
    $^7$Sterrenkundig Observatorium, Universiteit Gent, Krijgslaan 281 S9, 9000 Gent, Belgium\\
    $^8$Physics Department, Imperial College London, South Kensington Campus, London SW7 2AZ, UK\\
    $^9$Department of Physics and Astronomy, University of California, Irvine, CA 92697, USA\\
    $^{10}$INAF-Osservatorio Astronomico di Padova, Vicolo Osservatorio 5, I-35122 Padova, Italy\\
    $^{11}$SISSA, Via Bonomea 265, I-34136 Trieste, Italy\\
    $^{12}$Instituto de Fisica de Cantabria (CSIC-UC), Avda, los Castros s\slash n, 39005 Santander, Spain\\
    $^{13}$Leiden Observatory, Leiden University, P.O. Box 9513, NL - 2300 RA Leiden, the Netherlands\\
    $^{14}$Space Science and Technology Department, Rutherford Appleton Laboratory, Chilton, Didcot, Oxfordshire, OX11 0QX\\
    $^{15}$School of Physics and Astronomy, University of Nottingham, University Park, Nottingham, NG7 2RD, UK\\
    $^{16}$Institute of Astronomy, University of Cambridge, Madingley Road, CB3 0HA, UK\\
    $^{17}$UK Astronomy Technology Centre, The Royal Observatory, Blackford Hill, Edinburgh, EH9 3HJ, UK\\
    $^{18}$Pontifica Universidad Cat\'olica de Chile, Departamento de Astronom\'ia y Astrof\'isica, Vicu\~na Mackenna 4860, Casilla 306, Santiago 22, Chile
}
\begin{document}

\date{\today}

\pagerange{\pageref{firstpage}--\pageref{lastpage}} \pubyear{2013}

\maketitle

\label{firstpage}

\begin{abstract}
We use galaxies from the {\it Herschel}-ATLAS survey, and a suite of
ancillary simulations based on an isothermal dust model, to study our
ability to determine the effective dust temperature, luminosity and
emissivity index of 250\,$\mu$m selected galaxies in the local
Universe ($z < 0.5$). As well as simple far-infrared SED fitting of
individual galaxies based on $\chi^2$ minimisation, we attempt to
derive the best global isothermal properties of 13,826 galaxies with
reliable optical counterparts and spectroscopic redshifts. Using our
simulations, we highlight the fact that applying traditional SED
fitting techniques to noisy observational data in the {\it Herschel
  Space Observatory} bands introduces artificial anti-correlation
between derived values of dust temperature and emissivity index. This
is true even for galaxies with the most robust statistical detections
in our sample, making the results of such fitting difficult to
interpret. We apply a method to determine the best-fit global values
of isothermal effective temperature and emissivity index for $z < 0.5$
galaxies in \hatlas, deriving $\tgrey\ = 22.3 \pm 0.1$\,K and $\beta =
1.98 \pm 0.02$ (or $\tgrey\ = 23.5 \pm 0.1$\,K and $\beta = 1.82 \pm
0.02$ if we attempt to correct for bias by assuming that \teff\ and
\beff\ are independent and normally distributed). We use our technique
to test for an evolving emissivity index, finding only weak
evidence. The median dust luminosity of our sample is $\log_{10}
(\ldust \slash L_\odot) = 10.72 \pm 0.05$, which (unlike \teff) shows
little dependence on the choice of $\beta$ used in our analysis,
including whether it is variable or fixed. In addition, we use a
further suite of simulations based on a fixed emissivity index
isothermal model to emphasize the importance of the \hatlas\ PACS data
for deriving dust temperatures at these redshifts, even though they
are considerably less sensitive than the SPIRE data. Finally, we show
that the majority of galaxies detected by \hatlas\ are normal
star-forming galaxies, though with a substantial minority ($\sim
31$\,per cent) falling in the Luminous Infrared Galaxy category.
\end{abstract}


\section{Introduction}
\label{sec:intro}

Far-infrared spectral energy distributions (SEDs) of astrophysical
objects (both galactic and extra-galactic) are widely described
throughout the literature using an optically thin, single component
modified black-body emission profile as a function of frequency,
$\nu$, the so-called ``grey body'' model of the form in equation
\ref{eq:gb}.

\begin{equation}
f_\nu \propto
\frac{\nu^{3+\beta}}{\exp\left(\frac{h\nu}{kT}\right) -
  1}.
\label{eq:gb}
\end{equation}

\noindent Here, $T$ represents the isothermal temperature of the
source in question, $k$ is the Boltzmann constant and $h$ is the
Planck constant. This model is distinct from the traditional Planck
law since it accounts for a dust emissivity which varies as a
power-law with frequency, with the additional variable introduced
($\beta$ in equation \ref{eq:gb}) known as the emissivity index, which
varies between typical values of $1.5 < \beta < 2.0$. The value of
$\beta$ encodes information related to the dust grain properties
(e.g. grain composition and size\slash growth, details of the
absorption process) and may also vary with the temperature of the dust
\citep[see e.g.][]{hildebrand83,draine84,coupeaud11}.

Since far-infrared observations typically only sample the dust SED
sparsely (e.g. using observations at 60 and 100\,$\mu$m with {\it
  IRAS}, or 450 and 850\,$\mu$m using SCUBA), and given that equation
\ref{eq:gb} has three free parameters (\tgrey, $\beta$ and
normalisation), it has often been necessary to assume standard values
for the emissivity index to enable estimates of temperature and\slash
or dust luminosity to be constrained by the available data
\citep[e.g.][]{boulanger96,mcmahon99,smith08,smith09,smith10,bernard10,dye10,dunne11,orjales12}. This
is a practice which remains common despite the relatively good far-IR
SED sampling available using the PACS and SPIRE instruments aboard the
{\it Herschel Space Observatory} \citep[hereafter {\it
    Herschel}][]{pilbratt10}.

Observational constraints on the emissivity index using sparsely
sampled data are inevitably weak, and other studies have noted
anti-correlation between derived isothermal temperatures and
emissivity indices based on simple $\chi^2$ minimisation
\citep[e.g.][]{chapman03,dupac03,desert08,anderson10,paradis10,schnee10,veneziani10,bracco11,planck11,liang11,galametz12,paladini12}. As
some of these studies have noted, the interpretation of this effect is
difficult, especially given the near degeneracy between the two
parameters when observations only sample the Rayleigh-Jeans side of
the dust SED. In particular, a recent study by \citet{shetty09a},
noted that the best-fit values of \tgrey\ and $\beta$ appear
artificially anti-correlated using simple best-fit $\chi^2$ fits, even
in the hypothetical case where the two are intrinsically positively
correlated, due to the effects of noisy photometry on $\chi^2$ fitting
\citep[an effect which can be further exacerbated by having different
  temperature dust clouds along a line of sight,
  e.g. ][]{shetty09b,veneziani13}.

These effects were further demonstrated by \citet{kelly12}, who used
noisy model far-IR photometry in five {\it Herschel} bands to compare
simple $\chi^2$ best fit estimates for temperature and emissivity
index with values derived using a hierarchical Bayesian fitting
technique, with the latter method being able to recover the model's
intrinsic positive correlation between the two, in stark contrast to
the former which was hopelessly biased towards recovering
anti-correlation. \citet{kelly12} then proceeded to use their
technique on {\it Herschel} observations of the star-forming Bok
globule CB244 to show that temperature and emissivity index appear
weakly positively correlated in this Galactic source, though the
temperature baseline sampled was small, and the range of emissivity
indices at a given temperature was wide.

Previous studies of galaxies at far-infrared wavelengths ($\lambda >
50$\,$\mu$m) have noticed a correlation between the total dust
luminosity measured between 8 and 1000\,$\mu$m, \ldust, and the
effective temperature of the dust (so-called
``luminosity-temperature'' or ``L-T'' relations;
\citep[e.g.][]{kennicutt98,dunne00,dunne01,blain03,chapman03,coppin06,kovacs06,cde10,magnelli12,roseboom12,symeonidis13}. Though
it is probable that galaxies have multiple dust components of
different temperatures and emissivities superposed in the emergent
spectrum, an effective temperature and emissivity index -- hereafter
\tgrey\ and \beff\ -- can be useful in other areas, such as for
deriving rest-frame fluxes, calculating luminosities, dust-masses, or
luminosity\slash dust-mass functions
\citep[e.g.][]{dunne11,dye10,cortese12}, using the far-IR luminosity
as a star formation rate indicator based on monochromatic fluxes
\citep[e.g.][]{alejo09,smith08,smith09}, or using far-infrared
observations to derive approximate photometric redshifts for distant
sources \citep[e.g.][Pearson et al. {\it
    submitted}]{carilliyun00,amblard10}. This remains true, despite
the fact that these effective values are extremely difficult to
interpret in terms of the underlying properties of the interstellar
medium or stellar birth clouds in galaxies. Clearly, given the
possible dependence of temperature upon emissivity index mentioned
above, the possible influence of $\beta$ on any ``L-T'' relation is
great, potentially altering both the slope and normalisation (though
this would depend on the form, sign and strength of any intrinsic
temperature--emissivity-index relationship).

We discuss the \hatlas\ observations in section
\ref{sec:observations}, while we discuss the individual galaxy SED
fits used in our analysis in section \ref{sec:method}, and highlight
some problems resulting from allowing temperature and emissivity index
to vary using a traditional fitting technique based on $\chi^2$
minimisation. In section \ref{fitting:hier} we use an alternative
approach to determine the best global dust parametrisation for these
galaxies, under the simplifying assumptions that effective temperature
and emissivity index of the dust SED are independent, and that the
population is broadly homogeneous with normally distributed \teff\ and
\beff\ \citep[we note that the distributions of \teff\ and
  \beff\ derived in ][ are broadly consistent with being
  Gaussian]{kelly12,veneziani13}. In section \ref{sec:temprecovery} we
analyse the sensitivity of \hatlas-like photometry to different
effective dust temperatures and luminosities, assuming a fixed
emissivity index equal to our best estimate derived in section
\ref{fitting:hier}. We show that the temperature\slash luminosity
sensitivity of our technique as applied to the \hatlas\ data set shows
little bias with redshift, at least at $z < 0.5$ (where the vast
majority of the cross-identified galaxies in \hatlas\ lie). We
highlight the strong dependence of the derived results upon the
quality of the input photometry (particularly the availability of data
obtained using the PACS instrument), informing our understanding of
the fixed-$\beta$ properties of \hatlas\ galaxies in section
\ref{sec:fixedbeta_results}. Finally, we mention possible implications
for ``L-T'' relations suggested by other studies in the literature,
and make some concluding remarks in section \ref{sec:conclusions}. We
assume a standard cosmology with $H_0 = 71$\,km\,s$^{-1}$\,Mpc$^{-1}$,
$\Omega_M = 0.27$ and $\Omega_\Lambda = 0.73$ throughout.

\section{Observations}
\label{sec:observations}

We base our analyses on the phase 1 catalogue of the \hatlas\ survey
(Valiante et al., {\it in prep}), which consists of observations at
100 and 160\,$\mu$m from the PACS instrument \citep{poglitsch10} and
250, 350 and 500\,$\mu$m using the SPIRE instrument \citep{griffin10}
aboard the {\it Herschel Space Observatory}. This data release covers
$\sim$161.0\,deg$^2$ spread over the three equatorial regions of the
Galaxy And Mass Assembly survey \citep[GAMA - ][]{driver11}. Sources
brighter than $5\sigma$ in any single band were extracted from the
SPIRE data using the MAD-X algorithm (Maddox et al., {\it in prep})
applied to the maps made using the method descibed in
\citet{pascale11}. PACS fluxes were derived using apertures placed on
the maps \citep{ibar10} at the locations of the 250\,$\mu$m positions;
the \hatlas\ catalogue is described in detail in \citet{rigby11}. The
5\,$\sigma$ point source flux limits are 130, 130, 30.4, 36.9,
\&\ 40.8\,mJy in the 100, 160, 250, 350 \&\ 500\,$\mu$m bands,
respectively, with beam sizes ranging from 9 to 35\,arc sec FWHM in
the shortest- and longest-wavelength bands. Some details of the number
of formal detections in our sample are given in table
\ref{tab:sample}.

\begin{table}
\caption{The number of sources detected at $\ge 3$ \&\ $\ge 5\sigma$
  in each of the PACS and SPIRE bands in our sample, which consists of
  13,826 $\ge 5\sigma$ 250\,$\mu$m sources assigned reliable SDSS
  identifications with high-quality spectroscopic redshifts $z < 0.5$
  (see section \ref{sec:observations} for details). The SNR estimates
  are based on the ``BEST'' photometry values in the HATLAS catalogue,
  which are the values on which our fitting is based.}  \centerline{
\begin{tabular}{lcc}
Band & $N (\ge 3\sigma)$ & $N (\ge 5\sigma)$ \\
\hline
100\,$\mu$m & 2625 & 1122 \\
160\,$\mu$m & 3232 & 1235 \\
250\,$\mu$m &      & {\bf 13826} \\
350\,$\mu$m & 9141 & 3788 \\
500\,$\mu$m & 1922 & 522 
\end{tabular}}
\label{tab:sample}
\end{table}

In constructing our far-infrared catalogue we also include calibration
errors equal to 10 per cent of the magnitude of the flux for the PACS
bands, and 7 per cent for the SPIRE bands, by adding them in
quadrature to the estimated errors on the photometry. In the current
\hatlas\ release (Phase 1 - Valiante et al., {\it in prep}) we have
estimates of the PACS and SPIRE flux densities and their uncertainties
for every $5\sigma$ 250\,$\mu$m source\footnote{Every source {\it
    except} for 109 sources with SDSS r-band isophotal semi-major axis
  $> 30$\,arc sec in size, for which we cannot derive reliable fluxes
  at this time due to the high pass filtering used in the current PACS
  maps. We exclude these from our analysis, though we expect this
  issue to be fixed for the public release of the \hatlas\ data and
  refer the reader to \url{www.h-atlas.org} for the full technical
  details.} and it is these sources on which we base this analysis.

In order to derive redshifts for the $5\sigma$ 250\,$\mu$m sources, we
applied the same Likelihood Ratio (LR) algorithm discussed by
\citet{smith11} to derive reliable SDSS $r$-band counterparts to the
sources in the 250\,$\mu$m catalogue. In this analysis, we only
consider sources with reliable counterparts and spectroscopic
redshifts, which are mostly drawn from the SDSS \citep{york00} seventh
data release \citep{sdssdr7paper} and the GAMA survey
\citep{driver11}. The phase 1 catalogue contains 103,718 sources with
a signal-to-noise ratio of $> 5\sigma$ in the ``BEST'' photometry at
250\,$\mu$m. Of these, 29,053 have reliable matches to galaxies in the
SDSS $r$ band, with spectroscopic redshifts for 13,826 at $z < 0.5$
(approximately 48\,per cent of the SDSS-identified sample). We only
consider sources at $z < 0.5$ since at these redshifts the
\hatlas\ sample is dominated by the normal star-forming galaxy
population, while at higher redshifts active galactic nuclei
constitute an increasing fraction of the cross-identified 250\,$\mu$m
sources \citep{hardcastle10,serjeant10,bonfield11}.

\section{Traditional $\chi^2$ fitting}
\label{sec:method}

\subsection{Galaxy dust temperatures \& emissivity indices: individual fits}
\label{sec:gb_fitting}

For the 13,826 250\,$\mu$m sources with $z_{\mathrm{spec}} < 0.5$, we
derive isothermal grey-body fits of the standard form (equation
\ref{eq:gb}) on a galaxy-by-galaxy basis. We perform $\chi^2$
minimisation comparing each galaxy in our sample to a stochastic model
library, and in doing so build probability distribution functions
(hereafter PDFs\footnote{Technically speaking, these PDFs could
  instead be referred to as ``likelihood distribution functions'', but
  since several previous studies \citep[e.g.][as well as Kaviraj et
    al., {\it submitted}]{dacunha08,rowlands12,smith12} have used this
  terminology, we adopt it here for consistency.}) for each galaxy of
\tgrey, \beff\ and \ldust, by assuming that $P \propto
\exp\left(\frac{-\chi^2}{2}\right)$, and that each PDF is
marginalised. In this way we determine not only best-fit
(i.e. maximum-likelihood) values of each parameter for every galaxy,
but also median-likelihood values, with uncertainties derived
according to the 16th and 84th percentiles of the PDF. We also derive
two-dimensional PDFs showing the co-variance between the three
parameters (as determined by our fitting) in an analogous
manner. Since we have flux density estimates for every source in each
far-infrared band, we do not consider upper limits in our SED fitting.

Care must be taken when interpreting PDFs - whether they are for
individual galaxies, or stacks of a sample of galaxies - since they do
not represent the intrinsic properties of a source (or sample of
sources), but those values convolved with our ability to constrain
them. For the case of a single galaxy, with an intrinsic isothermal
dust SED, a PDF might be thought of as a delta function convolved with
(i.e. ``blurred'' by) the noise distribution for that particular
parameter; stacked PDFs instead represent the probability distribution
that would be {\it observed} if the sample could be drawn many times
from the observations and the analysis repeated. This is a point to
which we shall return in section \ref{sec:simulations}, below.

Rather than compute the transmission of greybody SEDs through the PACS
and SPIRE filter curves on the fly, we compute a stochastic library of
model photometry, binned in redshift at $\Delta z = 0.01$, spanning
$0.00 < z < 0.50$, which includes 50,000 versions of equation
\ref{eq:gb} in each redshift bin (allowing for the possibility that
sources in \hatlas\ may be extended by including photometry derived
using both the point- and extended-source SPIRE response curves). The
library assumes flat prior distributions of temperature and emissivity
index, ranging between $5 < T < 100$\,K and $-2.5 < \beta < 7.5$. We
highlight that we do not believe that such a wide range of values for
$\beta$ is physical, but we use it to illustrate the effects of
treating $\beta$ as a free parameter, and to avoid the PDFs being
truncated by the bounds of the prior where possible. Our method is
resistant to finding non-global minima \citep[e.g.][]{juvela12}, since
our flat priors stochastically sample the full parameter space,
including both global and local $\chi^2$ minima.

Unlike studies based on two-component far-IR SED models
\citep[e.g.][]{magnelli12}, we derive good fits with reduced $\chi^2 <
2.0$ to the majority of sources in our \hatlas\ sample using a
single-component isothermal model ($\sim$95 per cent at fixed $\beta$,
or 98 per cent using a variable $\beta$ model). This includes fitting
our observations in the 100\,$\mu$m band, though our PACS data are
less sensitive than our SPIRE data, and this is at least partly due to
the restframe wavelengths sampled at the redshifts of
\hatlas\ galaxies \citep[unlike studying $z \sim 2$ SMGs at
  100\,$\mu$m, which sample rest wavelengths $\sim$20-40\,$\mu$m, and
  necessitate a multiple-component SED model - e.g.][]{magnelli12}).

Despite deriving good fits to the \hatlas\ photometry with
an isothermal model, we do not suggest that the dust in
\hatlas\ galaxies is truly isothermal \citep[e.g.][]{dunne01}.

\begin{figure*}
%\centering
\subfigure[Stacked marginalised one-dimensional PDFs for \ldust, \tgrey, and emissivity index, $\beta$.]{\includegraphics[width=0.8\textwidth]{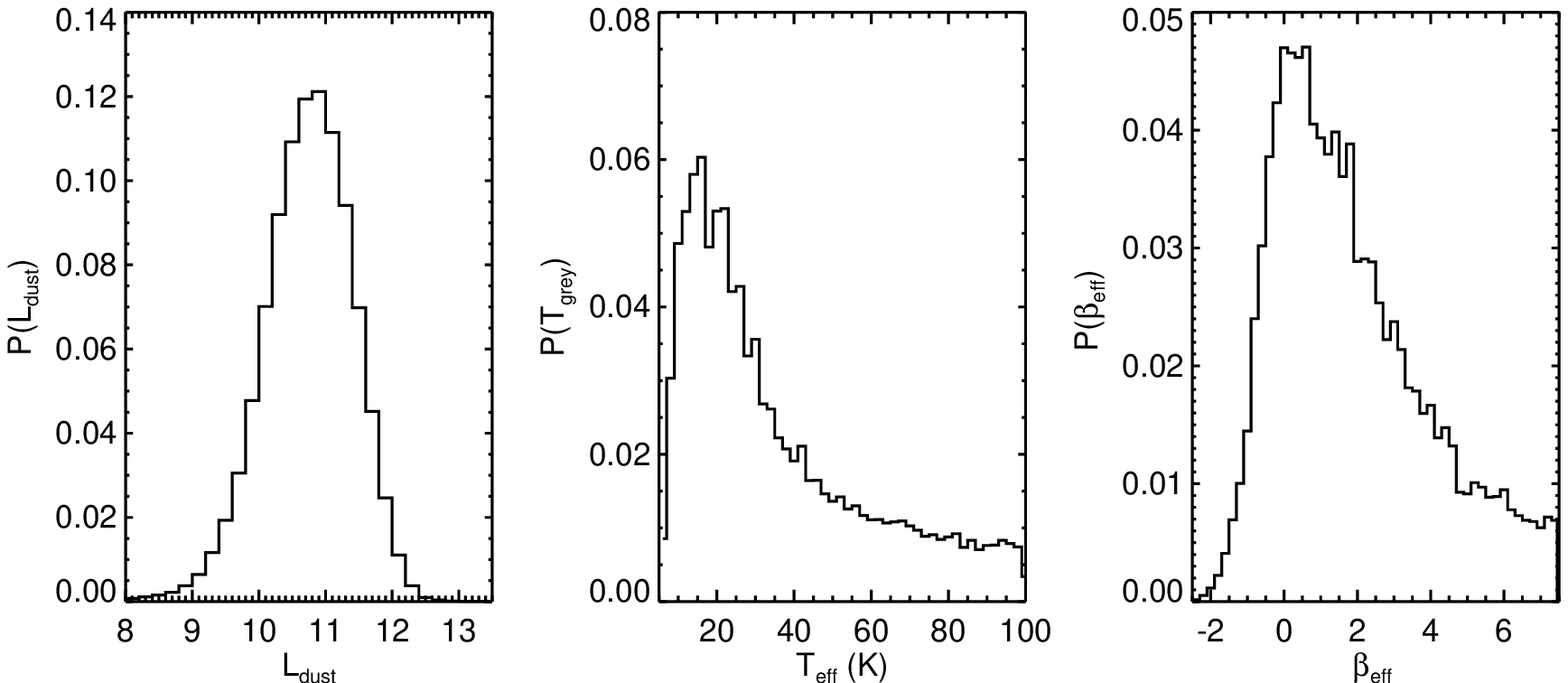}}
\subfigure[Two-dimensional stacked PDFs for the various combinations of \ldust, \tgrey\ and $\beta$. ]{\includegraphics[width=0.8\textwidth]{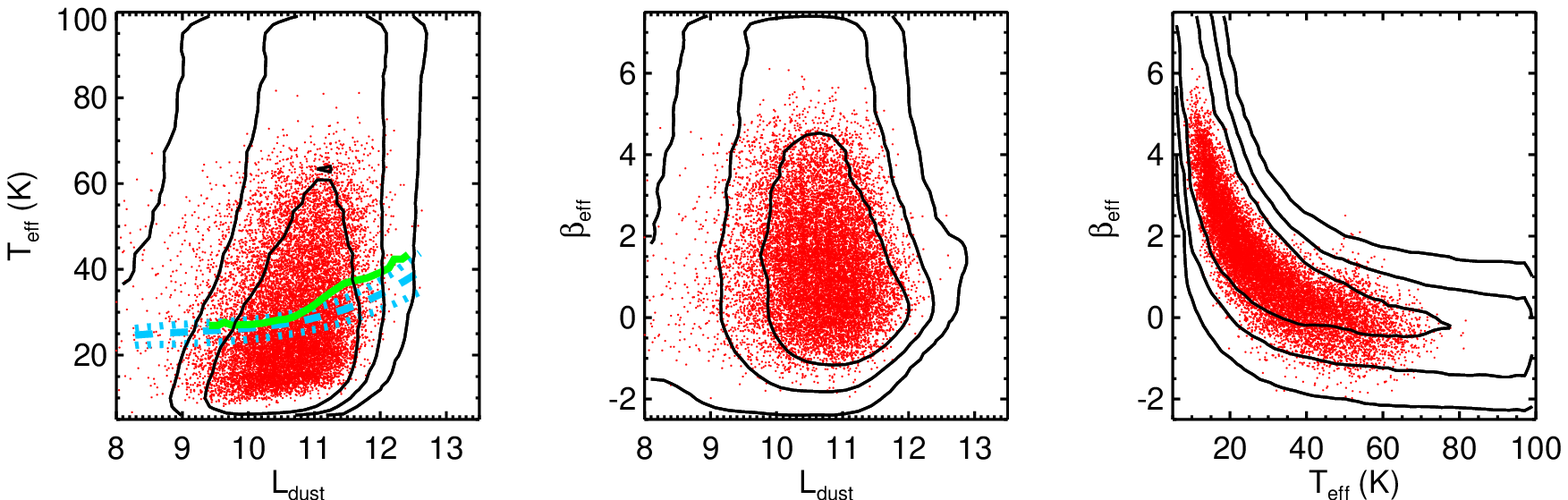}}
\caption{The (top) one- and (bottom) two-dimensional stacked PDFs for
  the galaxies in our sample which are well-described by the simple
  greybody model (i.e. those with reduced $\chi^2 < 2.0$). In the
  bottom panel, the median-likelihood esimates for each source are
  overlaid as the red points, while the ``L-T'' relations from
  \citet{chapman03} and \citet{hwang10} are shown as the light-blue
  and green lines, respectively.}
\label{fig:vb_results}
\end{figure*}

The results of our variable-$T$ and $\beta$ fitting are shown in
figure \ref{fig:vb_results}, with stacked\footnote{Stacked PDFs
  represent the sum of the individual PDFs for each galaxy\slash
  parameter of interest in a sample.} 1D PDFs for \ldust, \tgrey\ and
\beff\ in figure \ref{fig:vb_results}\,(a), and contour plots of the
stacked two-dimensional PDFs for the three different combinations of
parameters shown in figure \ref{fig:vb_results}\,(b), in which
median-likelihood values for each galaxy are overlaid in red. We also
overlay the dust luminosity-temperature relations from
\citet{chapman03}\footnote{This ``L-T'' relation has been corrected to
  account for the difference between L$_{\mathrm{FIR}}$ and
  L$_{\mathrm{TIR}}$ in \citet{chapman03}, assuming a redshift of
  $z=0.15$ and $\beta = 1.8$.} and \citet{hwang10} in light blue and
green, respectively.

Allowing both isothermal temperature and emissivity index to vary in
our fitting on a galaxy-by-galaxy basis gives several interesting
results. As figure \ref{fig:vb_results}\,(a) shows, the range of
values for \ldust\ appears quite reasonable, even when marginalising
over such a large range of temperatures and emissivities, while the
range of derived temperatures and emissivity indices is considerably
larger than can be found in the literature (despite the fact that we
do not explore the impact of alternative greybody SED parametrisations
on our results, and use the standard form given in equation
\ref{eq:gb} throughout). Though literature values for emissivity
indices typically vary between at most $1.0 < \beta < 2.5$, we find
that the range of values supported by the \hatlas\ data using this
method is extremely broad, though this \citep[in contrast to the
  sensible range of values for \ldust, consistent with that observed
  in \hatlas\ using full multiwavelength SED-fitting by
  e.g.][]{smith12} is hardly surprising given that the majority of
sources in our sample are only well-detected (i.e. $\ge 5\sigma$) in
the 250\,$\mu$m band \ref{tab:sample}. Interestingly, we also derive a
strong anti-correlation between \tgrey\ and \beff\ using this method,
in apparent agreement with the results of
\citep[e.g.][]{chapman03,desert08,anderson10,veneziani10,planck11}.

\begin{figure*}
\includegraphics[width=1.6\columnwidth]{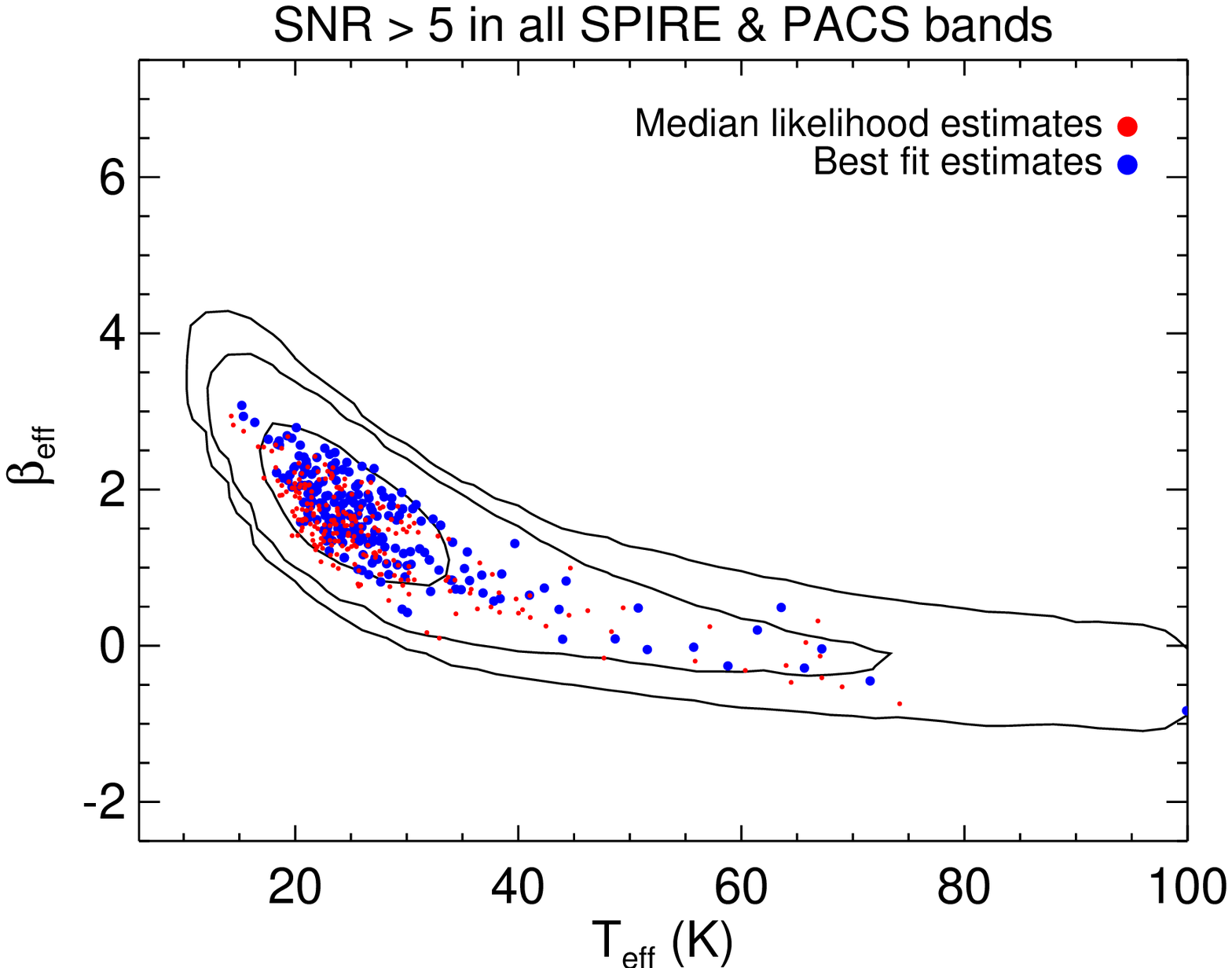}
\caption{The relationship between \teff\ and \beff\ for only those 238
  sources with the best photometry in \hatlas, with the best-fit
  (blue) and medium-likelihood (red) values overlaid on the 68.3,
  95.5, and 99.7 per cent confidence contours (i.e. equivalent to the
  1, 2, and $3\sigma$ levels assuming Gaussian statistics, in black).}
\label{fig:anticor_highsnr_real}
\end{figure*}

It is common practice to limit a sample to only include those sources
detected at the highest signal-to-noise ratio, in order to derive the
best constraints on parameters of interest. In \hatlas\ we have 238
galaxies which have photometry with signal to noise ratios $\ge 5.0$
in the ``BEST'' catalogue photometry in each of the PACS and SPIRE
bands. We show the results of applying our fitting technique to these
galaxies in figure \ref{fig:anticor_highsnr_real}, and once more
recover an apparently convincing anti-correlation between these two
parameters.

\subsection{Validation; simulations}
\label{sec:simulations}

\citet{shetty09a} suggested that an anti-correlation between
isothermal temperature and $\beta$ arises inevitably through the
influence of instrumental noise on simple $\chi^2$ fitting, and in a
partner publication \citep{shetty09b} through the presence of multiple
dust temperatures superposed along the line of sight \citep[see
  also][]{veneziani13}. To test whether the former result affects
\hatlas, we conducted a set of simple simulations designed to inform
our studies of \ldust, \teff\ and \beff\ based on the \hatlas\ data
set.

To populate our simulation, we first chose 200,000 temperatures,
randomly distributed between $11 < \tgrey\ < 56$\,K, and assigned each
temperature an emissivity index drawn at random from a Gaussian
distribution with median $\beta = 1.98$ and $\sigma (\beta) = 0.25$,
allowing us to generate a model intrinsic spectrum for each source.
We then computed the transmission of each model spectrum through the
PACS and SPIRE response functions to generate model noiseless
photometry in each of our photometric bands. Every model source in our
simulation now has an assigned temperature and emissivity index,
leaving only the normalisation of the model remaining to be
determined.

To avoid making any assumptions about the source number counts in
\hatlas, we assume that the observed fluxes of the model sources in
our catalogue (which in reality are a sum of the true flux from that
source, with additive noise based on the properties of the
observations in each band) are drawn from among the sources in the
\hatlas\ catalogue. We simulate the noise contribution to the measured
flux in each band differently for the PACS and SPIRE observations; for
PACS we generate a Gaussian distribution of noise values with standard
deviation equal to the 1$\sigma$ noise in the catalogues. In the SPIRE
bands, however, there is an additional contribution to the noise
component of the measured flux from source confusion \citep[which
  increases with wavelength;][]{nguyen10,rigby11}, making the true
noise distributions asymmetric.\footnote{The estimates for the
  1$\sigma$ confusion noise in the SPIRE bands are quoted by
  \citet{rigby11} to be 5.3, 6.4 and 6.7\,mJy beam$^{-1}$ at 250, 350
  and 500\,$\mu$m, respectively, compared with the 1$\sigma$
  instrumental noise values quoted in section \ref{sec:intro} of 3.0,
  3.7 and 4.7\,mJy beam$^{-1}$ in the same bands; the confusion noise
  is therefore larger than the instrumental noise contribution to the
  total noise in \hatlas\ SPIRE photometry.} To model the SPIRE noise
properties, we read off values at random positions from the
PSF-smoothed maps at 250, 350, and 500\,$\mu$m, automatically
including the instrumental and confusion noise components.

Using these model noise realisations, we determine the ``true''
250\,$\mu$m flux for each model source (i.e. the flux we would measure
in the absence of noise) and calculate the normalisation for each
model dust SED to the true 250\,$\mu$m flux density using equation
\ref{norm}:

\begin{equation}
\hat S_{250}^{\mathrm{model}} = \hat S_{250}^{\mathrm{obs}} -
\hat E_{250}^{\mathrm{model}}.
\label{norm}
\end{equation}

\noindent Where $\hat E_{250}^{\mathrm{model}}$ is the array of model
noise values measured from the 250\,$\mu$m maps, $\hat
S_{250}^{\mathrm{obs}}$ is the array of observed 250\,$\mu$m fluxes
drawn from \hatlas\ and $\hat S_{250}^{\mathrm{model}}$ is the precise
intrinsic flux of each model source. Calculating this ``noiseless''
250\,$\mu$m flux density for each source allows us to normalise each
model dust SED according to the noiseless value, and then by inverting
equation \ref{norm}, ``add back on'' the noise model for each of the
normalised PACS and SPIRE bands. In this way, we recover the observed
\hatlas\ 250\,$\mu$m flux density distribution exactly, and generate
model photometry in the other four PACS and SPIRE bands with realistic
noise characteristics. To ensure that our error estimates are
consistent with those in \hatlas, we assign an error on each noisy
model photometric data point by adopting the quoted error on the most
closely corresponding source in terms of flux density in the real
\hatlas\ catalogue on a band-by-band basis.

\begin{figure*}
\includegraphics[width=1.98\columnwidth]{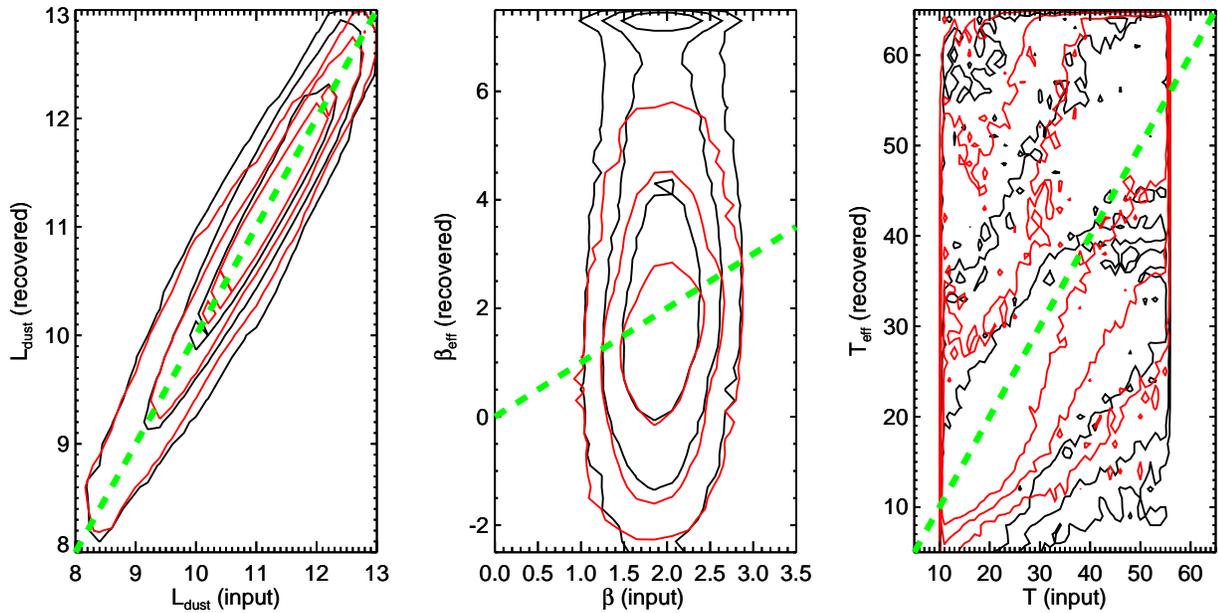}
\caption{Comparisons between the input values for \ldust, $\beta$ and
  \tgrey\ (left, centre and right-hand panels, respectively) and their
  median-likelihood (red) or best-fit (black) esimates in our
  \hatlas-like simulation. The contour levels correspond to the
  regions enclosing 68.3, 95.5 and 99.7 per cent of the recovered
  values. The green dashed lines indicate the ideal relation between
  input and recovered values, and the excess of best-fit sources
  around \beff\,(recovered) $\approx 7.5$ in the centre panel results
  from the tendency of best-fit values to accumulate where the
  posterior probability distribution is truncated by the bounds of the
  prior.}
% {FILE = SIM_REALCONF/COMPARE_INOUT.PRO}
\label{fig:vb_param_recovery}
\end{figure*}

In figure \ref{fig:vb_param_recovery} we show the variation between
the known input values in our simulation and the results of our
attempts to recover them. The red contours in figure
\ref{fig:vb_param_recovery} show the variation of the
median-likelihood estimates of \ldust, \beff\ and \tgrey, whilst the
black contours show the best-fit values (the contour levels represent
1, 2 \&\ 3$\sigma$ confidence levels in both cases). Though both
methods produce reasonable estimates of total dust luminosity, the
same can not be said of our attempts to recover either \beff\ or
\tgrey, which display large discrepancies from their known input
values, though the best-fit values show greater bias with larger
uncertainties than the median-likelihood values. This discrepancy from
the input values is in stark contrast to the results of section
\ref{sec:temprecovery}, in which we will derive temperatures and
luminosities at fixed emissivity index; clearly allowing $\beta$ to
vary has important implications for our results.

\begin{figure}
\includegraphics[width=0.98\columnwidth]{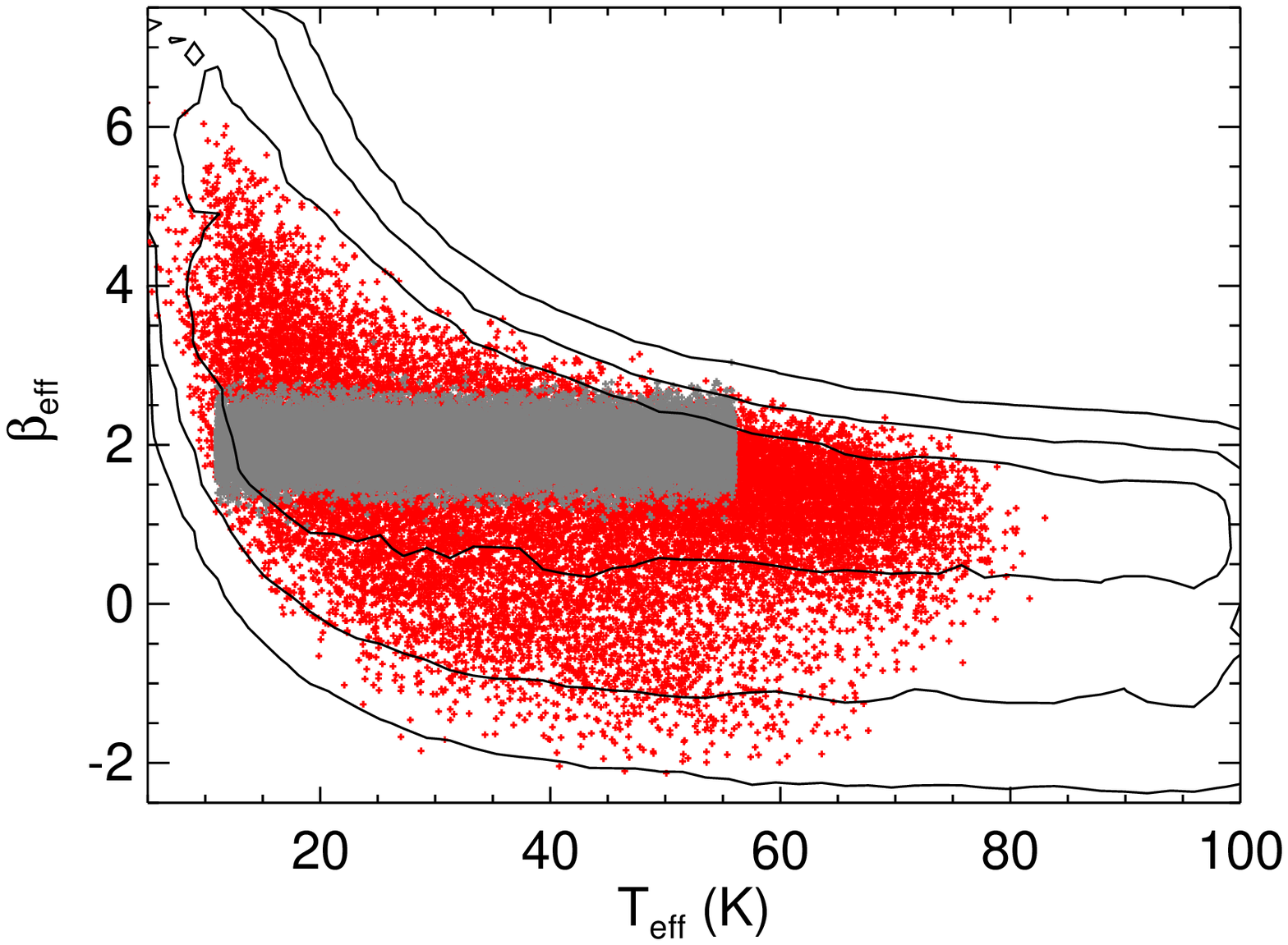}
\caption{Two dimensional stacked probability density function for
  \tgrey\ and $\beta$ shown by the solid black contours indicating the
  regions enclosing 68.3, 95.5 and 99.7 percent of the PDF. The
  median-likelihood estimates for each simulated galaxy well-described
  by our model are shown as the red points with the overlaid grey
  points indicating the input values, revealing the extent of the
  artificial anti-correlation.}
%{FILE = SIM_REALCONF/MAKE_ANTICOR_PLOT.PRO}
\label{fig:anticor}
\end{figure}

To investigate the large discrepancies between the input and output
temperatures and emissivity indices, in figure \ref{fig:anticor} we
show the relationship between \tgrey\ and \beff\ for the sources in
our simulation, with the input values in grey, the derived
median-likelihood values in red, and the contours representing the
stacked 2D PDF in black. Though the intrinsic relationship between
temperature and emissivity index in our simulation is flat, the noise
on our simulated photometry causes an artificial
temperature--emissivity index anti-correlation to be recovered. This
is in agreement with the results of \citet{shetty09a} and more
recently \citet{galametz12}, and explains the contrast between the
input and derived parameters in figure \ref{fig:vb_param_recovery}.

\begin{figure}
\includegraphics[width=0.98\columnwidth]{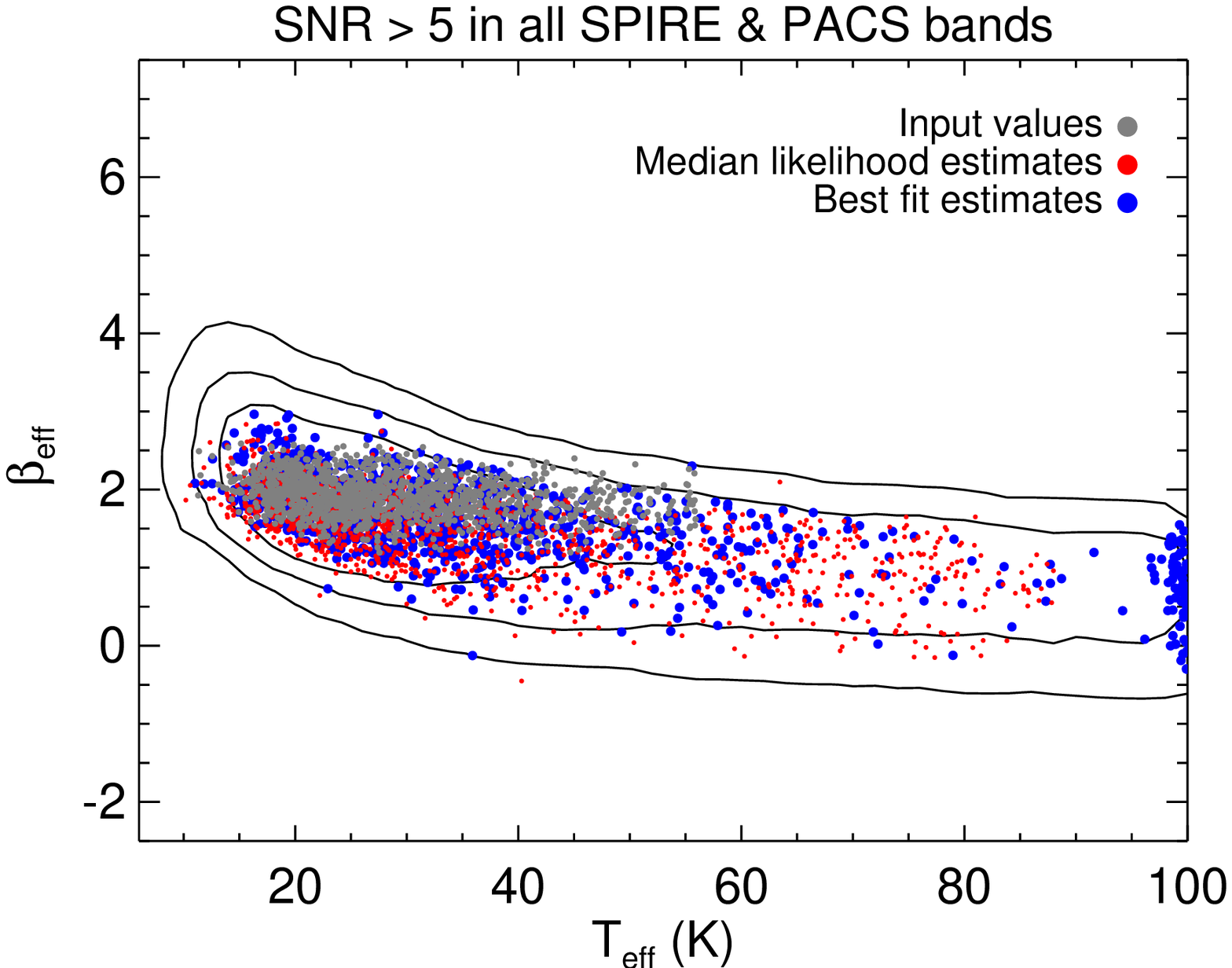}
\caption{Comparison between the values of temperature and emissivity
  index input to our simulation (grey points) and derived using
  best-fit (blue) and medium-likelihood (red) estimators. The clump of
  blue points reflects the tendency for best-fit values to be returned
  on the bounds of the temperature prior when the prior truncates the
  PDF. In contrast to figure \ref{fig:anticor}, this plot shows only
  sources with $\ge 5\sigma$ detections in each PACS\slash SPIRE
  band.}
\label{fig:anticor_highsnr}
\end{figure}

As mentioned in section \ref{sec:gb_fitting}, it is common practice to
limit a sample to include only those sources with the highest
signal-to-noise ratios in their photometry. We now use our simulation
to test whether this approach can mitigate the influence of the
anti-correlation. Figure \ref{fig:anticor_highsnr} shows the
relationship between the derived temperature and emissivity index
including only those simulated sources with $\ge 5\sigma$ detections
in each of the PACS and SPIRE bands. Once more the input values are
shown in grey, while the best fit values are shown in blue, with the
median-likelihood estimates in red. Though the range of recovered
\beff\ is unsurprisingly smaller than for the full sample, the
artificial anticorrelation still persists, though the range of
temperatures input to the simulation is broader than is observed in
\hatlas\ (our input distribution of model temperatures is flat between
$11 \le \teff \le 56$\,K).

\begin{figure}
\includegraphics[width=0.98\columnwidth]{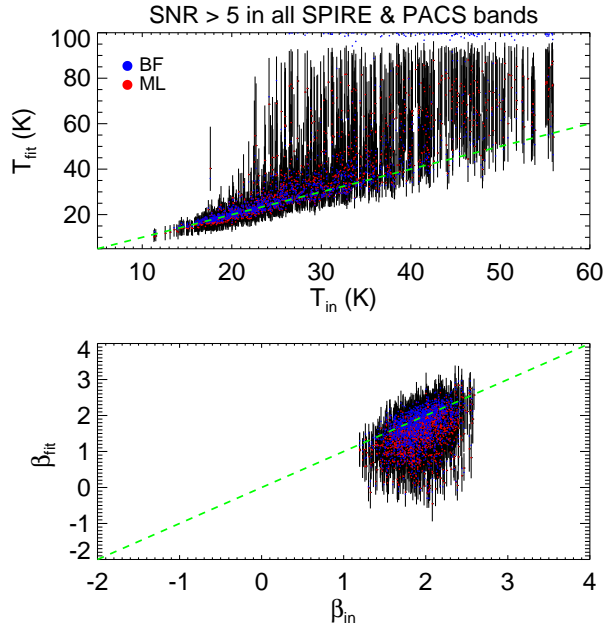}
\caption{A comparison between the input and derived temperatures (top)
  and emissivity indices (bottom), for simulated sources with $\ge
  5\sigma$ detections in each of the PACS\slash SPIRE bands, with
  unity highlighted using the dashed green line. The red circles with
  errors represent the median-likelihood estimates, while the blue
  points represent the best-fit values. It is clear that when both
  \teff\ and \beff\ are allowed to vary, temperatures are inevitably
  biased high, and emissivity indices show considerable negative
  bias.}
\label{fig:compare_io_highsnr}
\end{figure}

To highlight the bias, in figure \ref{fig:compare_io_highsnr} we show
the relationship between the individual recovered \teff\ and
\beff\ and their corresponding input values. The median likelihood
values are shown as the red circles with the black error bars
corresponding to the 16th and 84th percentiles of the individual PDFs,
while the best-fit values are shown as the blue circles. The recovered
temperatures show considerable bias above their input values
(represented by the dashed green line) and the emissivity indices are
biased low even when considering only the galaxies with the highest
SNR in our simulated PACS/SPIRE photometry.

\begin{figure*}
\includegraphics[width=1.98\columnwidth]{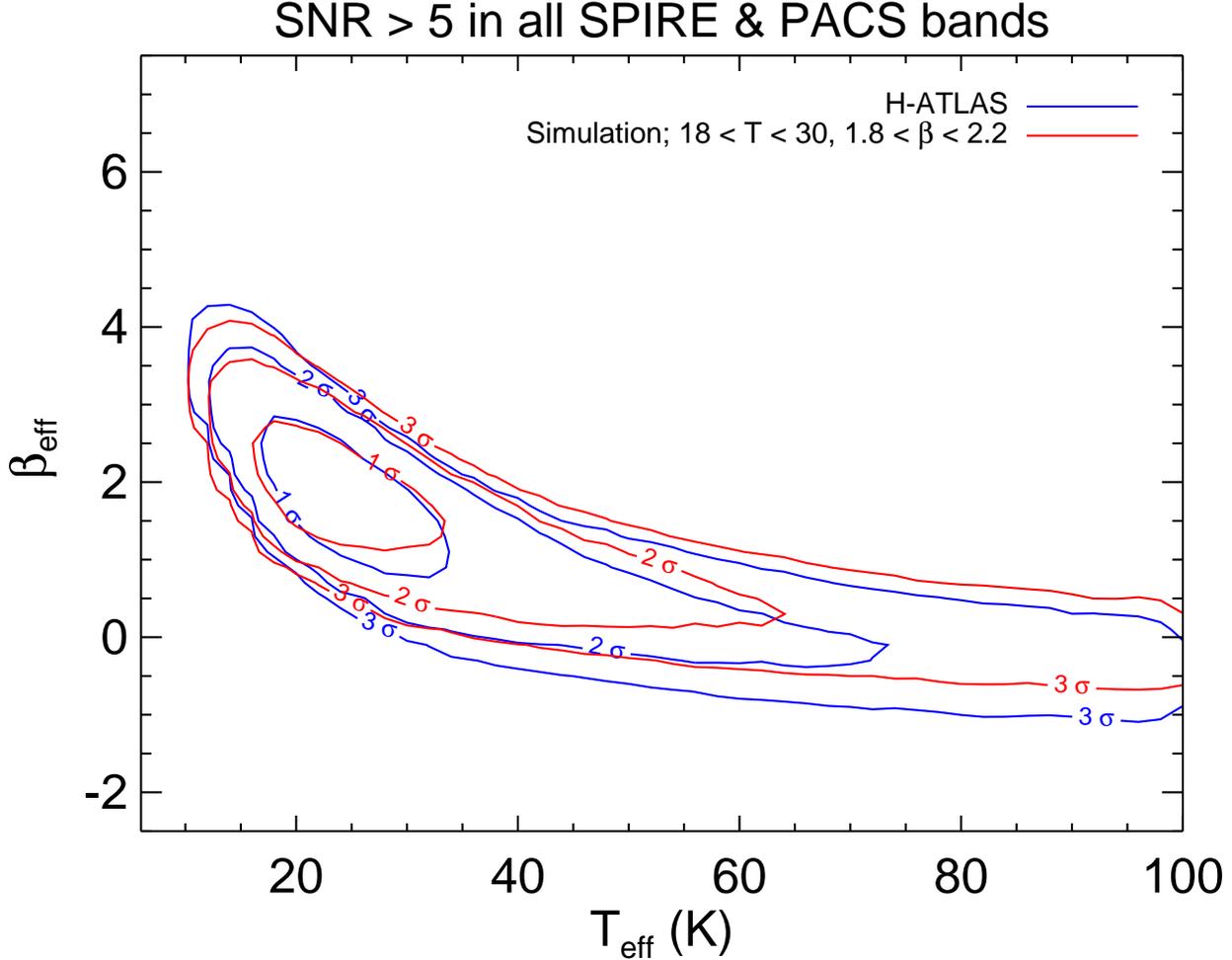}
\caption{Comparison between the 2D confidence intervals for \teff\ and
  \beff\ shown for the subset of the \hatlas\ catalogue detected at
  $\ge 5\sigma$ in all PACS\slash SPIRE bands (in blue) and for a
  similar subset of our simulation (in red), with the input values of
  temperature and emissivity truncated relative to the full
  simulation, as discussed in the text. The values input to our
  simulation for \teff\ and \beff\ have no intrinsic correlation.}
\label{fig:anticor_highsnr_sim}
\end{figure*}

Finally, if we reduce the range of the input values for temperature
and emissivity index that have gone in to our model sample, such that
$18.0 \le T \le 30.0$ and the Gaussian distribution of emissivity
indices is truncated at $1.8 \le \beta \le 2.2$, and again select only
those sources with $\ge 5\sigma$ detections in each {\it Herschel}
band, we recover results very similar to those recovered using the
corresponding subsample (in SNR) of the real \hatlas\ data set,
despite there being no intrinsic correlation between $\beta$ and
$T$. Figure \ref{fig:anticor_highsnr_sim} shows the 1, 2, and
$3\sigma$ confidence intervals of the 2D PDF for the highest-SNR
subset of \hatlas\ (in blue) and of our simulation (in red).

Since we have shown that we recover an apparent anti-correlation
between temperature and emissivity index, even though the two are
independent in the simulated data and despite having included only
those sources with the best-constrained SEDs, it is clear that any
evidence for correlation between \teff\ and \beff\ derived using this
method is weak.

\section{What are the parameters of the average isothermal \hatlas\ galaxy SED?}
\label{fitting:hier}

\subsection{Method}

As we mentioned in the introduction, use of standard values for
emissivity index and temperature of model far-infrared isothermal SEDs
is common at all redshifts; with this in mind, it is desirable to
determine our best global estimates of the temperature and emissivity
index for galaxies in \hatlas\ assuming this model.

Clearly, the artificial anti-correlation produced by the $\chi^2$
fitting precludes producing useful results on a galaxy-by-galaxy basis
using this technique. We use a simple method, similar to that used by
\citet{hardcastle13}, to determine the best-fit global \teff\ and
\beff. To do this, we treat the galaxies in \hatlas\ as a homogeneous
sample, and record the best-fit $\chi^2$ value in bins of
\beff\ allowing \tgrey\ and \ldust\ to vary (and similarly in bins of
\tgrey\ allowing \beff\ and \ldust\ to vary). We may then calculate
the sum of the $\chi^2$ values in each bin across the good fits in our
sample (i.e. those galaxies with $\chi^2_{\mathrm{best}} < 2.0$). This
technique has the advantage that by summing in $\chi^2$, those
galaxies with only weak constraints from the photometry have
approximately flat distributions of $\chi^2 (\teff)$ and
$\chi^2(\beff)$, while those galaxies with the best constraints
exhibit clear minima in these distributions, which when combined over
the whole sample produce strongly preferred values of \teff\ and
\beff. Furthermore, this approach not only increases the
signal-to-noise ratio by combining together the photometry for all of
the sources in a sensible way, but also naturally accounts for the
asymmetric error distributions for each individual source, which
preclude using a traditional weighted mean approach. Stacking the
values of $\sum\limits_i \chi_i^2 (\teff)$ and $\sum\limits_i
\chi_i^2(\beff)$ results in a naturally weighted distribution allowing
us to derive our best estimates of the two parameters.

Our model library contains 50,000 stochastic samplings of \beff\ and
\teff\ in each redshift bin (recall that our library samples redshift
at $\Delta z = 0.01$ for $0.00 < z < 0.50$); in order that our
histograms of $\chi^2(\teff)$ and $\chi^2(\beff)$ are
smoothly-varying, we use a resolution of $\Delta \teff = 2.0$\,K, and
$\Delta \beff = 0.2$. Since the sampling in temperature and emissivity
index in our histograms is relatively coarse, we assume that the
underlying distributions of $\chi^2$ are also smoothly varying, and
locate the minimum of each distribution by interpolating the five data
points about the minimum using a fourth order polynomial. We may then
use the polynomial fit to the total $\chi^2$ distribution to generate
PDFs for \beff\ and \teff\ in our sample by assuming the same
relationship between probability and $\chi^2$ as before. We derive
uncertainties using the 16th and 84th percentiles of the resulting
PDF. Due to the large number of sources in the sample, and the large
values of $\chi^2$ in each bin, the derived errors are smaller than
the histogram bins, reflecting the need to interpolate between
them. Finally, in contrast to the results for individual galaxies, the
maximum- and median-likelihood values that we derive using this method
are very similar since the PDFs obtained from the polynomial fits to
the histograms of $\sum\limits_i \chi^2_i$ are very well constrained.

\begin{figure*}
\includegraphics[width=1.95\columnwidth]{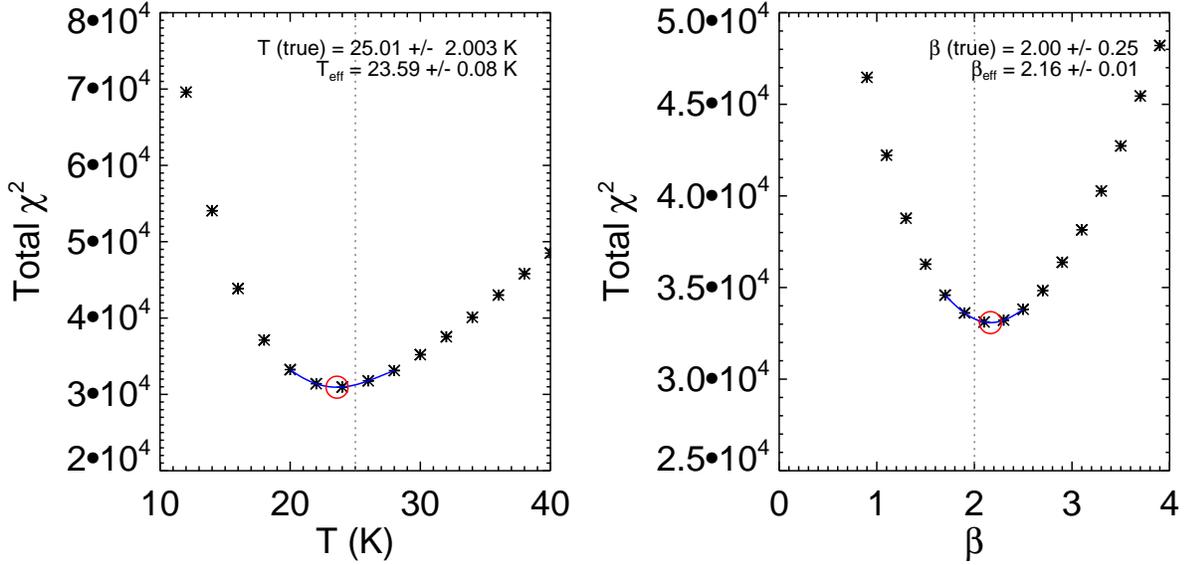}
  \caption{Plots illustrating the method for recovering known values
    of \tgrey\ and $\beta$ using our technique. We sample $\beta$
    (left) and temperature (right) and calculate the total $\chi^2$
    over the whole sample in each bin. We model the $\chi^2$
    distribution about the bin with the lowest value, using a fourth
    order polynomial (in blue), and find the minimum value by
    interpolating along this polynomial and assuming that the
    distribution of values in $\chi^2$ space is smooth. The determined
    minimum is shown by the red circles in each plot, with the true
    values for this particular library ($\beta = 2.00 \pm 0.25$ and
    $\tgrey = 25.0 \pm 2.0$) indicated by the vertical dashed grey
    lines.}
  \label{fig:hier_method}
\end{figure*}

In figure \ref{fig:hier_method} we show examples of the results of
using this method to determine our best estimates of the population
mean \tgrey\ and \beff, to compare the results with the known input
values from a similar simulation to the one discussed in section
\ref{sec:simulations}. However, since we are now interested in the
global properties of galaxies in \hatlas\ rather than determining the
bias in our fitting techniques, we assume input values to our
simulation which are Gaussian-distributed about $\beta = 2.00 \pm
0.25$ and $\tgrey = 25.0 \pm 2.0$, uncorrelated with $\beta$. The
left-hand panel of figure \ref{fig:hier_method} shows typical results
derived using $\sim$14,000 model galaxies described using the
isothermal model. Though using this technique limits the influence of
the aforementioned artificial anti-correlation introduced through
line-of-sight effects and noisy photometry, the derived values for
$\beta$ (left panel) still show residual systematic bias towards
higher values, while \tgrey\ (right) still shows slight bias towards
colder temperatures than the inputs.

\begin{figure*}
  \centering 
  \includegraphics[height=0.98\columnwidth]{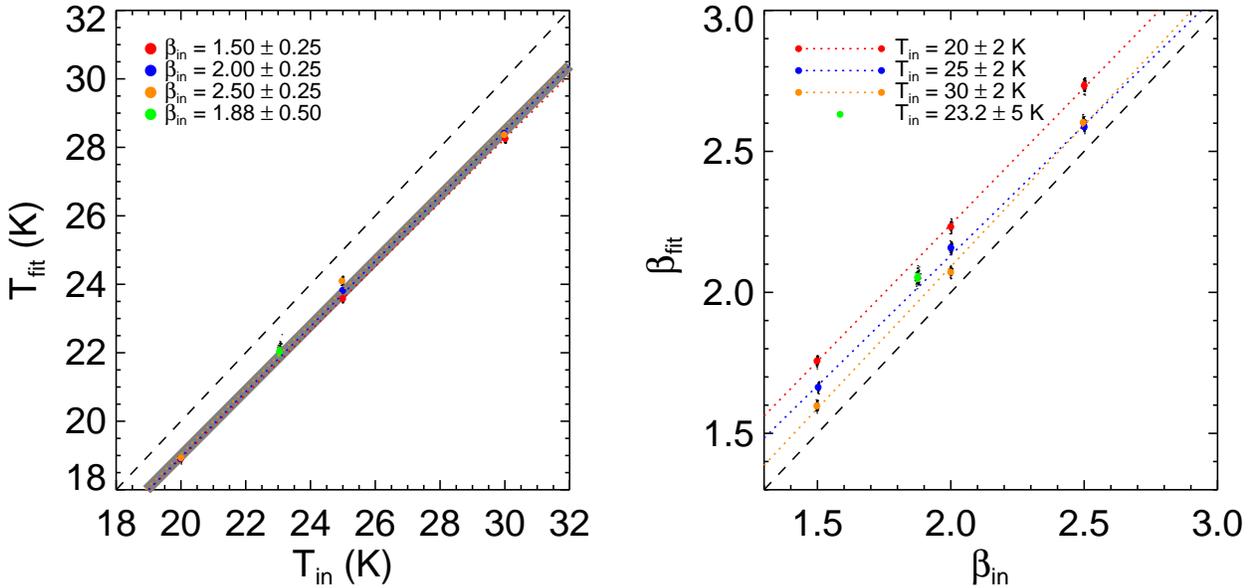}
  \caption{Recovering temperatures (left) and $\beta$ values (right)
    using our technique applied to 100 realisations of ten
    combinations of temperature and emissivity index. The input values
    are spaced on a grid of \tgrey\ and $\beta$, with \tgrey\ = 15, 20
    and 25\,K and $\beta = 1.5$, 2.0 and 2.5 as shown by the coloured
    points (see legend), and each simulation assumes that temperature
    and emissivity index are independent and Gaussian distributed; the
    derived errors in the recovered values are smaller than the
    plotting symbols. We also include a broader distribution of input
    temperatures and emissivities centred about $\tgrey = 23.18 \pm
    6.1$ and $\beta = 1.88 \pm 0.5$ to demonstrate the validity of our
    approach for both narrow and wide input distributions. The dashed
    grey line shows the ideal relation (i.e. perfect recovery of each
    parameter) in both the left and the right panels. The thick grey
    line in the left panel shows the best fit linear relationship
    between the input and output temperatures as detailed in equation
    \ref{eq:tgrey_correction}. The dashed coloured lines in the
    right-hand plot (for the recovery of $\beta$) show the best-fit
    linear relationship between the input and derived values in bins
    of input temperature. We correct for this offset as a function of
    temperature using equation \ref{eq:beta_correction}, as described
    in the text.}
  \label{fig:hier_validation}
\end{figure*}

To illustrate this method, and better quantify the bias, we show the
results of recovering the known input values on 100 Monte-Carlo
realisations of a grid of Gaussian distributed temperatures about 15,
20 and 25 (all $\pm\ 2$\,K), and uncorrelated $\beta = 1.5$, 2.0 and
2.5 (all $\pm 0.25$) in figure \ref{fig:hier_validation}. We also
include another input sample arbitrarily centred on $\tgrey = 23.2 \pm
5.0$\,K and $\beff = 1.88 \pm 0.50$ to show that the results are
reasonable for broad as well as narrow distributions of \tgrey\ and
\beff. The results of recovering \tgrey\ are shown in the left panels,
while the results for \beff\ are shown on the right. The values of
\tgrey\ recovered by our fitting of this simulation
($T_{\mathrm{fit}}$) are consistently biased lower than the input
values ($T_{\mathrm{in}}$), with the best fit linear relationship
(thick grey line) biased below the ideal relationship (dashed grey).

The best-fit linear relationship between input and recovered
temperatures (the solid line in figure \ref{fig:hier_validation}),
under the assumptions that \tgrey\ and \beff\ are independent, and
Gaussian distributed about some mean value, is given by equation
\ref{eq:tgrey_correction}:

\begin{equation}
  T_{\mathrm{fit}} = (0.949 \pm 0.001)T_{\mathrm{input}} + (0.023 \pm 0.039).
  \label{eq:tgrey_correction}
\end{equation}

\noindent This relationship does not show any strong dependence on the
value of $\beta$ in the simulation, reflected by the coloured points
and error bars in the left-hand panels in figure
\ref{fig:hier_validation} which are all consistent with equation
\ref{eq:tgrey_correction} (the thick grey line).

\begin{figure}
  \centering \includegraphics[trim=0cm 0cm 0cm 10cm, clip=true,
    width=0.98\columnwidth]{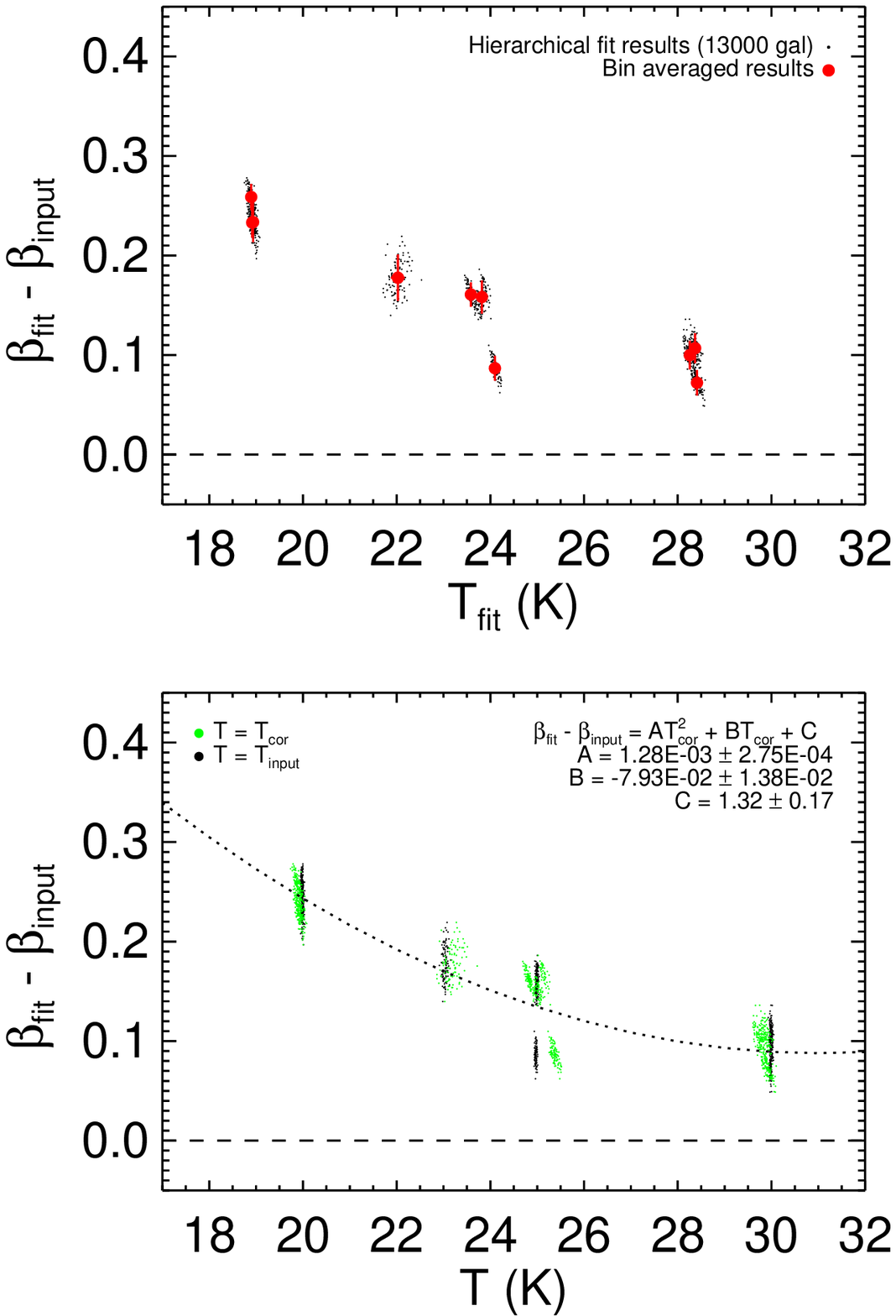}
  \caption{The difference between the values for $\beta$ inserted in
    to and recovered from one hundred realisations of our simulations,
    as a function of the recovered ``corrected'' temperature (in
    green) and the true (i.e. input) temperature (in black). The best
    2nd order polynomial fit to $\beta_{in} - \beta_{fit}$ as a
    function of corrected recovered temperature is overlaid as the
    dotted grey line, with the fit parameters detailed in the legend,
    and in equation \ref{eq:beta_correction}.}
  \label{fig:hier_betacorr}
\end{figure}

Our results for recovering $\beta$ are more complicated, with the bias
to higher emissivity indices being more pronounced for the colder
sources on the model grid, and the corresponding offsets between the
coloured dotted lines (the best-fit linear relationships for each
input temperature bin). We show the temperature dependence of the
$\beta$-bias in figure \ref{fig:hier_betacorr}, with each point
representing one of 100 Monte-Carlo realisations sampling 14,000 model
galaxies. We show the bias as a function of input temperature in
black, and as a function of the recovered temperatures, after
correcting for the fitting offset using equation
\ref{eq:tgrey_correction}, overlaid in green. To quantify the bias in
$\beta$ as a function of \tgrey, we derive a 2nd-order polynomial fit
between the corrected temperatures and $(\beta_{\mathrm{fit}} -
\beta_{\mathrm{input}})$, with the values for each parameter shown in
the upper right-hand corner of figure \ref{fig:hier_betacorr}. The
best fit 2nd-order polynomial is given by equation
\ref{eq:beta_correction}:

\begin{align}
  \begin{split}
  \beta_{\mathrm{fit}} - \beta_{\mathrm{input}} =\  &(1.28 \pm 0.28) \times 10^{-3}\ T_{\mathrm{cor}}^2\\
  &- (7.93 \pm 1.38)\times 10^{-2}\ T_{\mathrm{cor}} \\
  &+ (1.32 \pm 0.17).
  \end{split}
  \label{eq:beta_correction}
\end{align}

\noindent As a crude test of whether combining the galaxies in this
way limits the impact of anti-correlation on our results as compared
with a more traditional fitting approach, we compare the degree of
anti-correlation recovered by the two methods.

Figure \ref{fig:hier_betacorr} reveals a change of $\Delta \beta
\approx 0.15$ between $T=20 -- 30$\,K ($\Delta\beta \slash \Delta T
\approx 0.015$), whilst the traditional $\chi^2$ fitting (figure
\ref{fig:anticor}) suggests a change of $\Delta \beta \approx 1$ over
a similar range of temperature ($\Delta \beta \slash \Delta T \approx
0.1$). These values suggest a factor of $> 6$ times improvement in
terms of the influence of the anti-correlation (though at the expense
of estimates of \teff\ and \beff\ for the individual sources). This is
perhaps even more notable given that we do not account for confusion
noise in our SED fitting.

We assume uncertainties on the corrected temperatures and emissivity
indices equal to the standard deviation of the corrected values
(derived using equations \ref{eq:tgrey_correction}
\&\ \ref{eq:beta_correction}) about their medians, added in quadrature
to the mean of the 16th and 84th percentiles of the derived PDFs.

It is tempting to use these offsets based on our simulations to
crudely attempt to correct for the bias in the derived values of
\teff, however, these offsets are only really valid for the particular
case in which \beff\ is independent of \teff; to test the strength of
this assumption, we generated additional suites of simulations in
which \teff\ and \beff\ have plausible intrinsic correlation or
anti-correlation\footnote{The [anti-]correlations that we insert are
  linear relations assuming that $\beta \propto AT$, allowing $A = \pm
  0.16$. To avoid unbounded $\beta$ for high temperatures and
  unphysically low $\beta$ for low temperature, we fixed $\beta$ at
  the boundary values for $T < 20$ and $T>28$\,K.}. When we attempt to
recover the input values using our method on the correlated
simulations, the offsets we recover are $\Delta \teff =
T_{\mathrm{recovered}} - T_{\mathrm{true}} = -1.51 \pm 0.11$\,K and
$-0.71 \pm 0.13$ for the anti-correlated and correlated simulations,
respectively, as compared with $\Delta \teff = -1.14$ assuming
equation \ref{eq:tgrey_correction}, suggesting that our temperature
corrections are reasonable to within $\sim 0.4$\,K. For $\Delta \beff$
(defined in the analogous manner), the values we recover are $0.00 \pm
0.02$ and $0.16 \pm 0.02$, as compared with $\Delta \beff = 0.18$
estimated using equation \ref{eq:beta_correction}.

Since we have shown that previous studies inevitably produced
artificially anti-correlated values of \teff\ and \beff\ using this
technique, and the best estimate of the variation between \teff\ and
\beff\ not subject to this limitation finds only weak positive
correlation between the two \citep[][albeit over a very small
  temperature range, and with a large range of $\beta$ at any given
  temperature; see the right-hand panel of their figure 6]{kelly12},
we proceed under the assumption that the two are effectively
un-correlated in \hatlas, though in what follows we quote both the raw
and the crudely ``corrected'' values.

\subsection{Global dust properties in \hatlas}
\label{sec:dust_properties}

\begin{figure*}
\centering
\includegraphics[width=1.90\columnwidth]{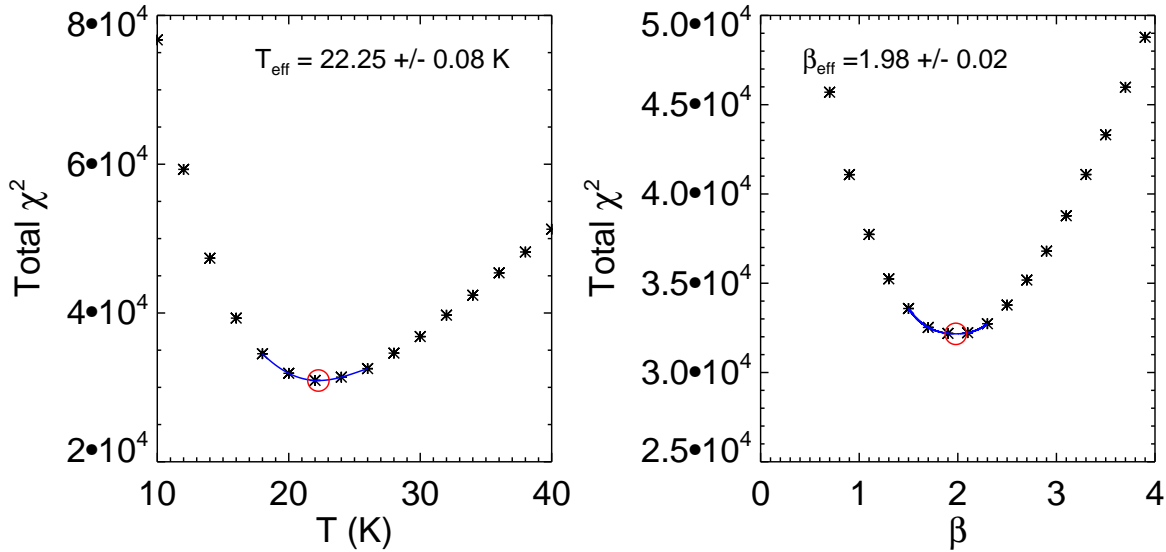}
\caption{SED fitting results for emissivity index $\beta$ and
  \tgrey\ in \hatlas, showing the distributions of $\sum \chi^2
  (\beta)$ and $\sum \chi^2 (T)$. The data are shown as black
  asterisks, while the best-fit fourth-order polynomial is shown as
  the blue line, with the location of the minimum (i.e. the best-fit)
  shown by the red circle, with the uncorrected values for each
  minimum quoted in the legend. The similarity of these values to
  those derived for our simulation in figure \ref{fig:hier_method}
  indicate the fidelity of our simulation to \hatlas.}
\label{fig:hier_res_hatlas}
\end{figure*}

We applied our method to the stacked $\chi^2$ distributions of the
12,814 out of 13,826 galaxies with best-fit reduced $\chi^2 <
2.0$. The results are shown in figure \ref{fig:hier_res_hatlas}; the
median likelihood values for 250\,$\mu$m-selected galaxies are $\tgrey
= 22.3 \pm 0.1$\,K and $\beff = 1.98 \pm 0.02$. If we use equations
\ref{eq:tgrey_correction} \&\ \ref{eq:beta_correction} to correct
these values in the same way as before (again highlighting that these
corrections assume independent \teff\ and \beff, and that the true
values are normally distributed) we derive our best estimates for
galaxies in \hatlas\ of $\tgrey^{\mathrm{corr}} = 23.5 \pm 0.1$\,K and
$\beff^{\mathrm{corr}} = 1.82 \pm 0.02$. Our estimate of $\beta$
compares well with the results of \citet{planck11}, who suggest a
distribution centred about $\beta = 1.78$ (albeit with ``significant
T-$\beta$ anti-correlation''). It is also consistent with the ranges
of values suggested by
\citet{chapin09,paradis10,bracco11,liang11,galametz12,magnelli12} and
\citet{roseboom13} for a wide range of sources, including galactic
star forming regions, galactic cirrus, resolved nearby galaxies and
high redshift sub-millimetre galaxies; that the values for $\beta$ are
consistent over such a wide variety of scales is remarkable.

%chapin = 1.75
%roseboom13 1.6+/-0.5
%magnelli 2.0+/-0.2

\begin{figure*}
\centering
\includegraphics[width=1.6\columnwidth]{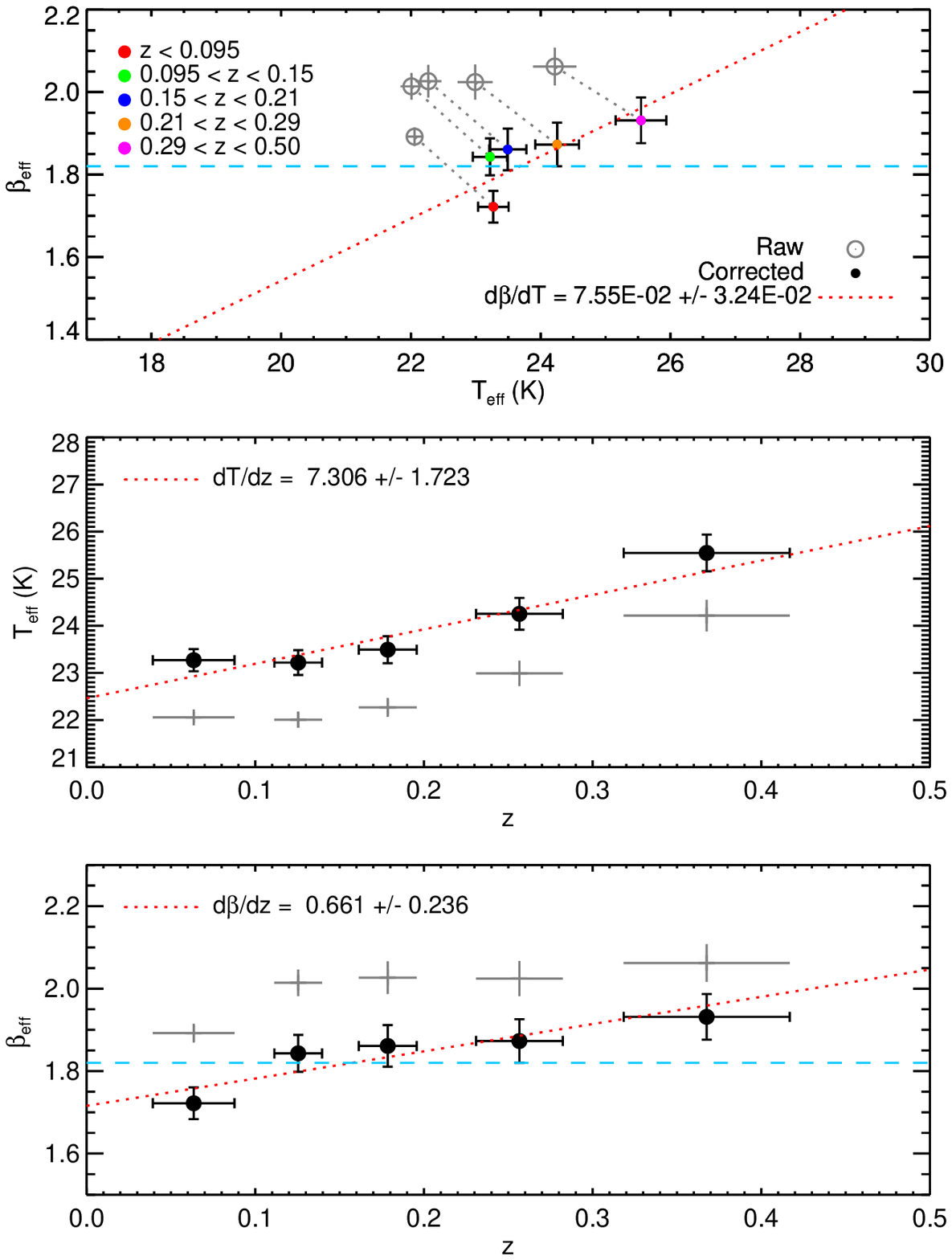}
\caption{Top: Comparison between the values derived using our method
  for \tgrey\ and $\beta$ in bins of redshift, such that there are
  approximately equal numbers of galaxies in each bin. Middle:
  Variation between the recovered \tgrey\ as a function of redshift,
  showing a $>4\sigma$ positive correlation. Bottom: recovered $\beta$
  as a function of redshift, showing only weak evidence for redshift
  dependance of emissivity index in \hatlas. The median $\log_{10}
  \left ( \ldust \slash L_\odot \right)$ for each redshift bin is
  10.01, 10.42, 10.67, 10.97 and 11.31, from the lowest to the highest
  redshift bin, respectively.}
\label{fig:z_t_beta_variation}
\end{figure*}

In figure \ref{fig:z_t_beta_variation} we show the results of applying
our fitting method once more, this time in bins of redshift, to test
for the possibility of variation in emissivity index or \tgrey. A
suite of simulations, similar to those discussed in section
\ref{sec:simulations} only with bins containing 2,600 objects, were
found to have \teff\ and \beff\ offset from their input values in a
manner consistent with equations \ref{eq:tgrey_correction} and
\ref{eq:beta_correction} derived using 14,000 objects per bin (albeit
with larger uncertainties). We use the same corrections for these
values, and update the uncertainties accordingly.

The bounds of the redshift bins, as well as the gradient of the
best-fit linear fit to each combination of parameters (in orange) are
detailed in the panel legends. In the top panel, we show that there is
only weak evidence for variation in \beff\ as a function of
\tgrey\ and of redshift (or dust luminosity; the median dust
luminosity in each redshift bin is given in the caption to figure
\ref{fig:z_t_beta_variation}), with the gradient of the linear best
fit relationship distinct from zero only at the $\sim2.3\,\sigma$
level. This large uncertainty reflects the possibility that the
relationship may be largely driven by a single outlying bin (perhaps
either the highest or lowest redshift bin); it is quite possible that
the highest redshift bin is biased towards higher temperatures due to
the higher average dust luminosity of the sources in this bin
(i.e. Malmquist bias) combined with the aforementioned ``L-T''
relation. In the middle plot we show the redshift variation of \tgrey,
showing a weak trend as expected from the ``$L-T$'' relations
suggested in previous works and overlaid in figure
\ref{fig:vb_results}; the gradient is significant at the 4.2\,$\sigma$
level. The median dust luminosity in each redshift bin also increases
with redshift, as expected. In the bottom panel, we compare the
derived emissivity indices with the best-fit value across the whole
sample (dashed light-blue line), again showing only weak evidence for
any variation (the gradient of the best-fit line, shown in dotted
orange, is formally only significant at the 2.8\,$\sigma$ level).

Though we have attempted to correct for residual bias in our results
using equations \ref{eq:tgrey_correction} and
\ref{eq:beta_correction}, these corrections are invalid if the two are
related (a hypothesis for which there is only weak statistical
evidence at the time of writing, though our simulations in section
\ref{fitting:hier} suggest that the additional uncertainty added to
the corrections in this case is only $\sim 0.4$\,K or a change in
emissivity of around 0.18). We leave further investigation of any
putative intrinsic relation between \teff\ and \beff\ for a future
investigation.

\section{Recovering temperature and luminosity at fixed emissivity index using \hatlas}
\label{sec:temprecovery}

Using isothermal models to describe the far-IR properties of
\hatlas\ sources holds the distinct advantage of enabling us to
compare dust temperatures with pre-{\it Herschel} studies. As we
explained in the previous sections, it is common for emissivity index
to be held fixed, since effective temperatures can be useful even with
a small number of data points available to constrain the far-infrared
SED. In what follows, we assume a fixed emissivity index, $\beta =
1.82$ corresponding to our best estimate of the global mean in
\hatlas, after correcting for bias as discussed in the previous
section. We will discuss the implications of the choice of $\beta$ in
section \ref{sec:fixedbeta_results}.

We perform $\chi^2$ minimisation comparing each galaxy in our sample
to a library of model photometry, based on dust SEDs derived using
equation \ref{eq:gb}, evaluated on a grid of temperatures between $5.0
< \tgrey\ < 65$\,K at 0.2\,K resolution. Once more, we account for the
transmission of each dust spectrum through the PACS and SPIRE response
functions, and build \tgrey\ and \ldust\ PDFs for each galaxy in the
same manner as before. We calculate best-fit values, and use the PDFs
to determine median-likelihood estimates of \tgrey\ and \ldust\ as
well as uncertainties, in the same manner as before.

To test the ability of our fitting to recover \teff\ and \ldust\ of
galaxies in \hatlas, we also built a fixed $\beta= 1.82$ version of
the \hatlas-like simulation mentioned above, and generated three
additional closely-related simulations:

\begin{itemize}
  \item We re-ran the simulation neglecting the SPIRE confusion noise
    from the modelling, instead using a symmetric Gaussian model noise
    distribution rather than the asymmetric combination of
    instrumental and confusion noise that blights the real data. The
    standard deviation of the symmetric model noise distribution is
    defined to be equal to the sum in quadrature of the instrumental
    and confusion noise distributions in the real \hatlas\ data set
    (i.e. the values quoted in section \ref{sec:observations}).
  \item We repeated simulating the model photometry, this time
    assuming that the PACS data have the same sensitivity as the SPIRE
    250\,$\mu$m data, to compare our results with those available
    using the greater sensitivity available using other, smaller area
    {\it Herschel} surveys, in particular the combination of the PACS
    Evolutionary Probe \citep[PEP;][]{lutz11} and the {\it Herschel}
    Multi-tiered Extragalactic Survey \citep[$Her$MES;
    ][]{oliver12}. We note that the only difference between this
    simulation and the ``\hatlas-like'' simulation (previous item in
    this list) is the sensitivity of the PACS data, not the observed
    source counts.
  \item Finally, we generated a simulation neglecting the PACS data,
    to test our ability to recover temperatures using the SPIRE
    photometry alone.
\end{itemize}

\noindent The results of fitting to these different simulated datasets
are discussed in the following subsections. The flat input
distribution of temperatures in our simulation enables us to test the
ability of our fixed $\beta$ $\chi^2$ fitting to recover galaxies
across a large range of temperature and dust luminosity. It is not
intended to be consistent with the \hatlas\ selection function, merely
to enable us to study the biases inherent in using different data sets
to study these parameters.

\subsection{Recovering known temperature and luminosity}
\label{subsec:recovery}

\begin{figure*}
\centering \subfigure[``Deep PACS''
  simulation]{\includegraphics[width=1.30\columnwidth]{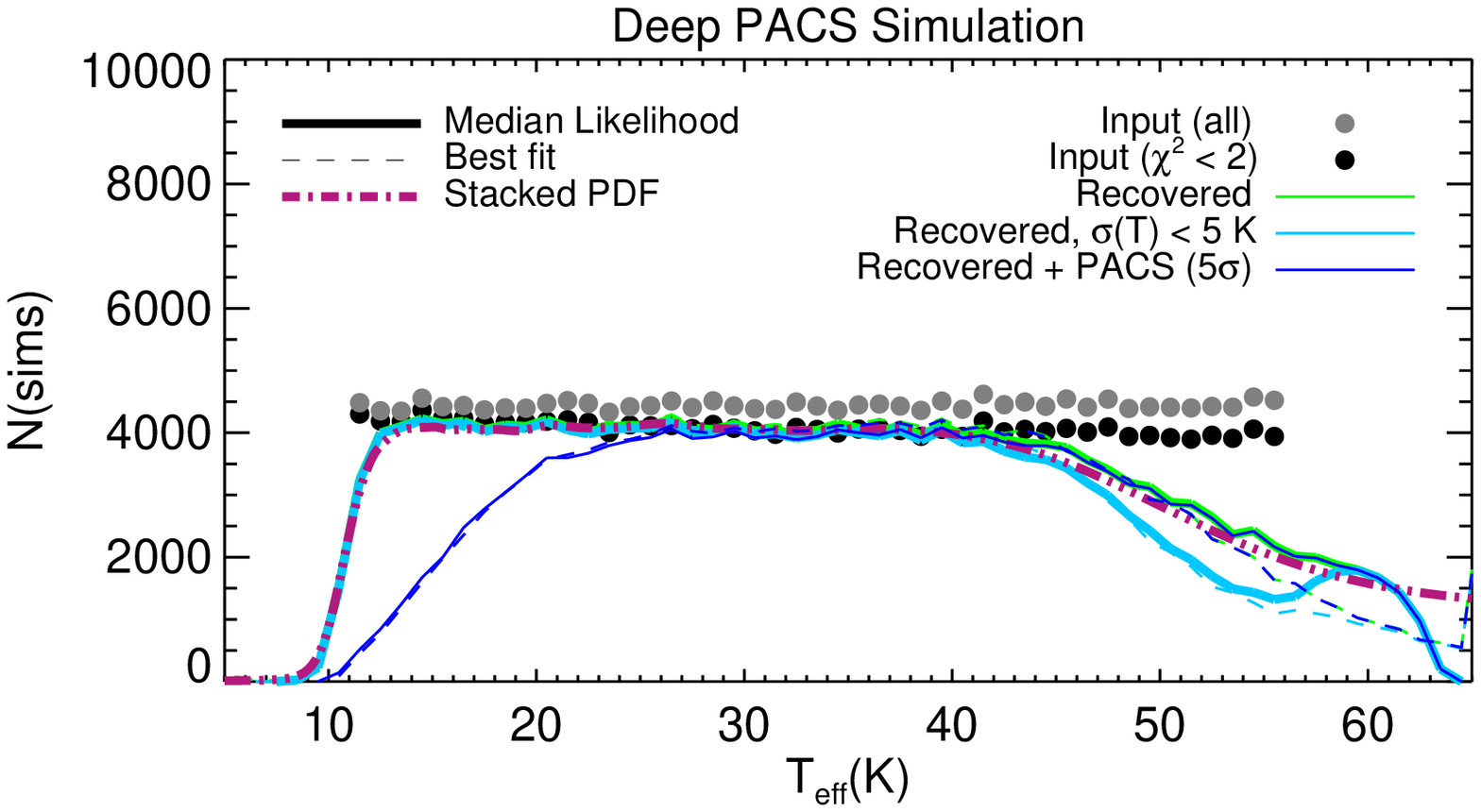}}
\subfigure[``\hatlas-like''
  simulation]{\includegraphics[width=1.30\columnwidth]{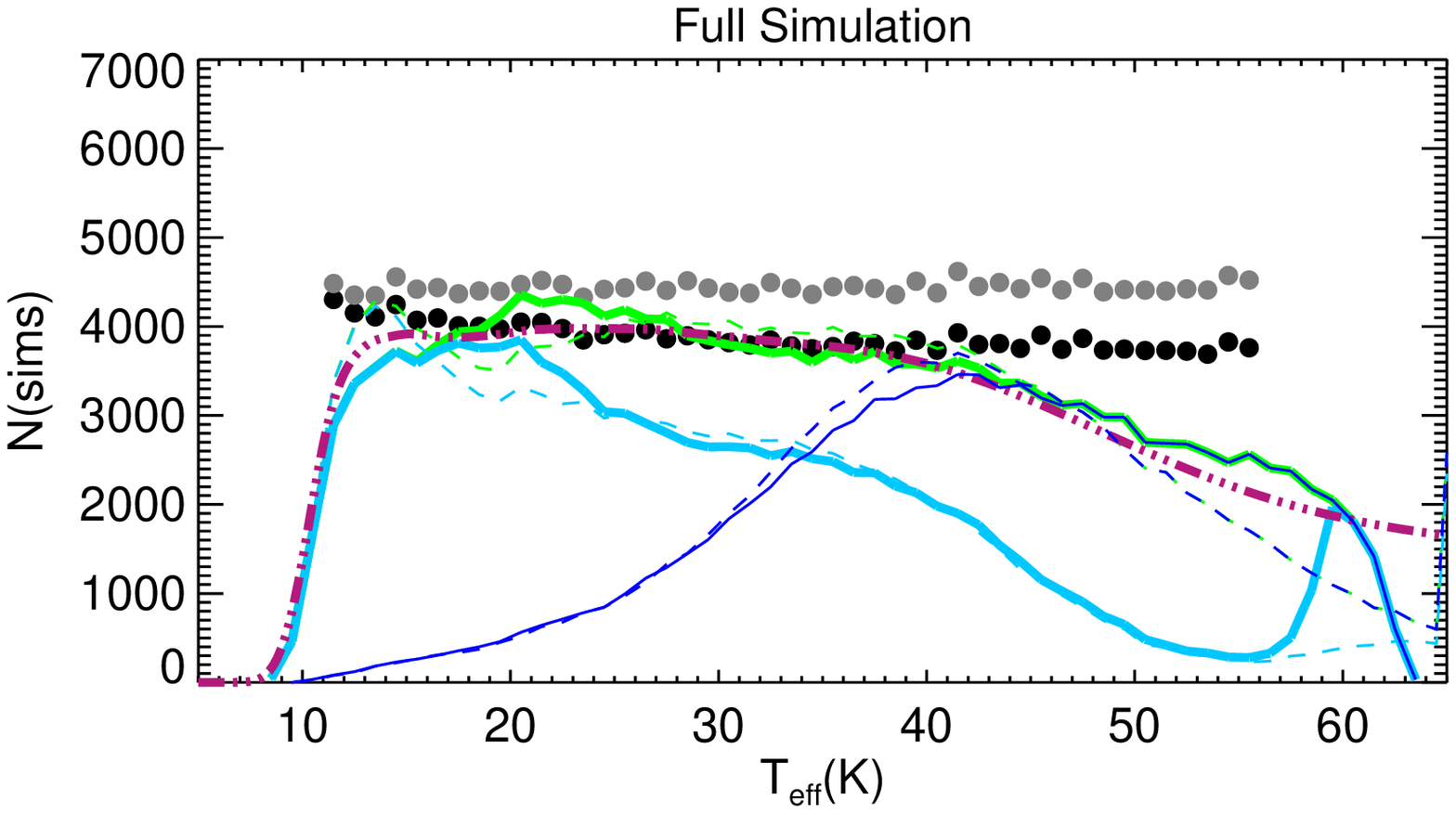}}
\subfigure[``No PACS''
  simulation]{\includegraphics[width=1.30\columnwidth]{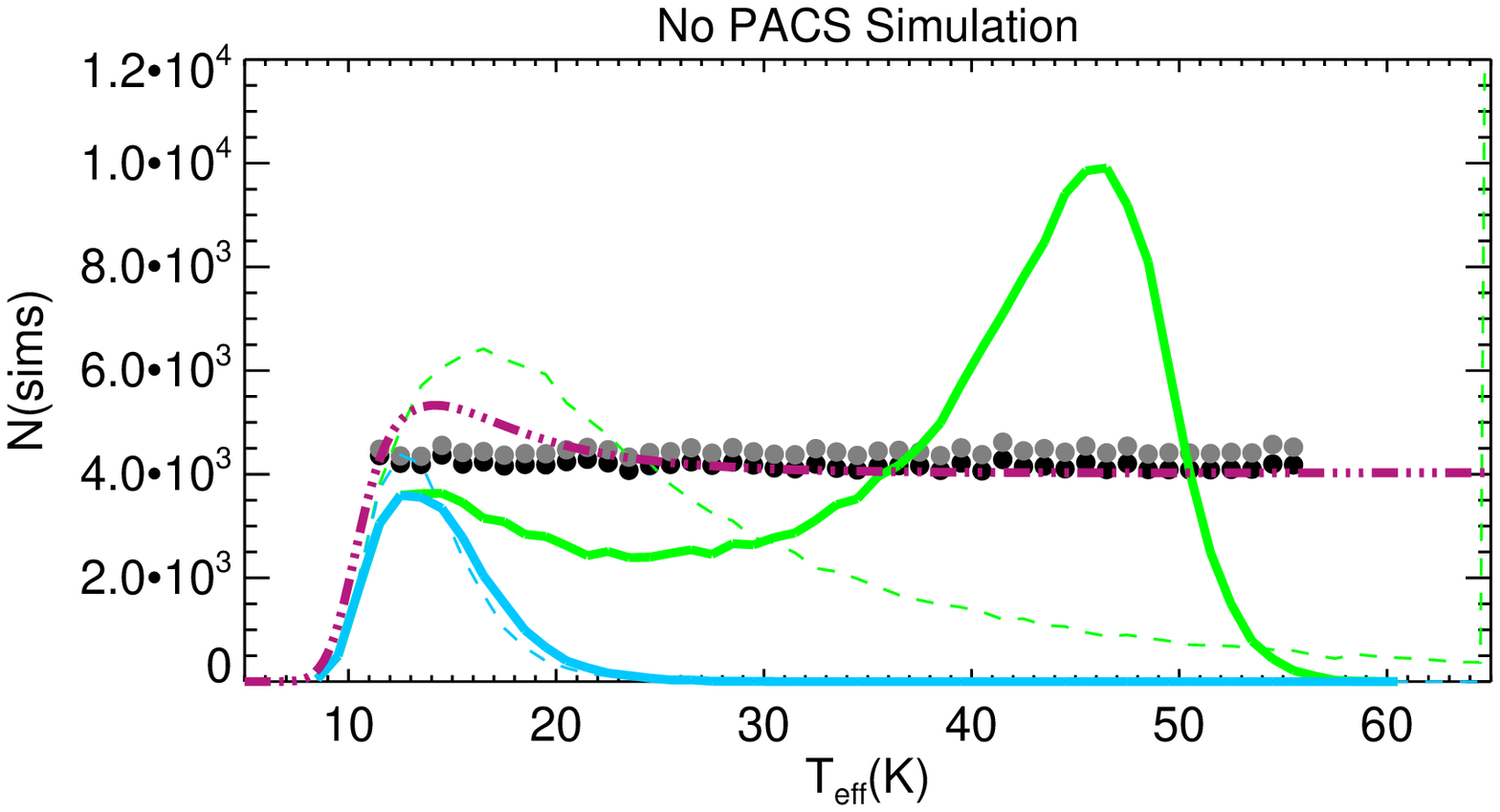}}
\caption{Histograms showing the input and output values of \tgrey\ in
  our fixed $\beta = 1.82$ simulations, ``Deep PACS'',
  ``\hatlas-like'' and ``No PACS'', from top to bottom, respectively,
  with the input values (i.e. true temperatures) indicated by grey
  circles. The distribution of input temperatures for the sources for
  which we derive good fits is shown in black, and the output results
  (i.e. the values recovered by our fitting, with no SNR cuts apart
  from the 250\,$\mu$m selection criterion) for the same set of
  galaxies is in green, allowing the two to be directly compared. We
  also include the distribution of the subsample of galaxies with good
  fits and good temperature constraints (i.e. $\sigma(\tgrey ) <
  5$\,K) in light blue. The histogram of recovered values for sources
  with $> 5\sigma$ PACS detections is shown by the darker blue lines
  (``recovered $+$ PACS'' in the legend). The median-likelihood
  recovered values are indicated by the solid lines, while the
  best-fit values are shown by the dashed lines of the same colour
  scheme, while the purple dot-dot-dashed lines show the renormalised
  stacked temperature PDF for all the galaxies with good fits ($\chi^2
  < 2.0$). }
\label{fig:tin_tout_hists}
\end{figure*}

\begin{figure*}
%  \vspace{18cm}
  \subfigure[``Deep PACS'' simulation]{\includegraphics[width=0.33\textwidth]{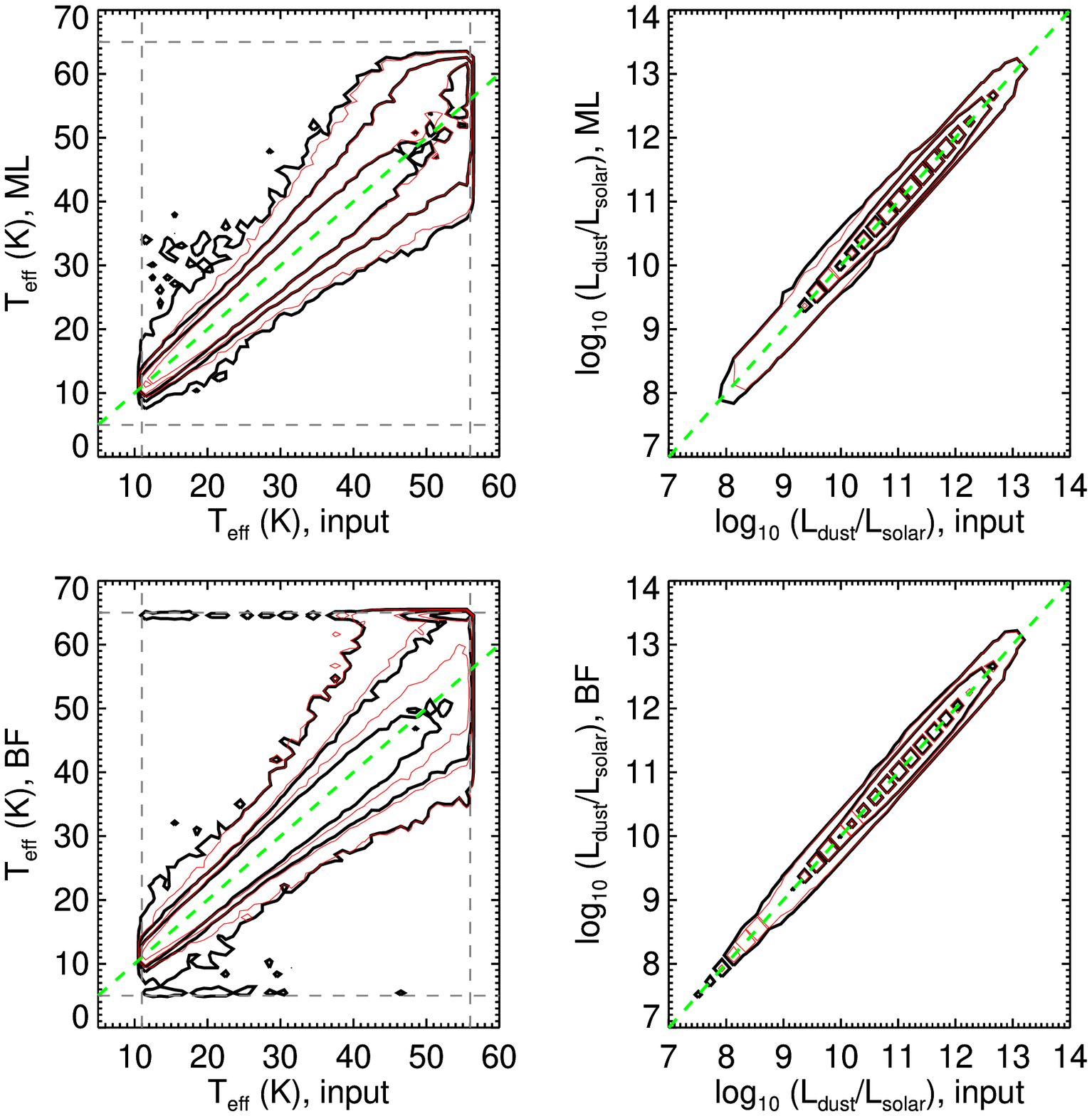}}
  \subfigure[``\hatlas -like'' simulation]{\includegraphics[width=0.33\textwidth]{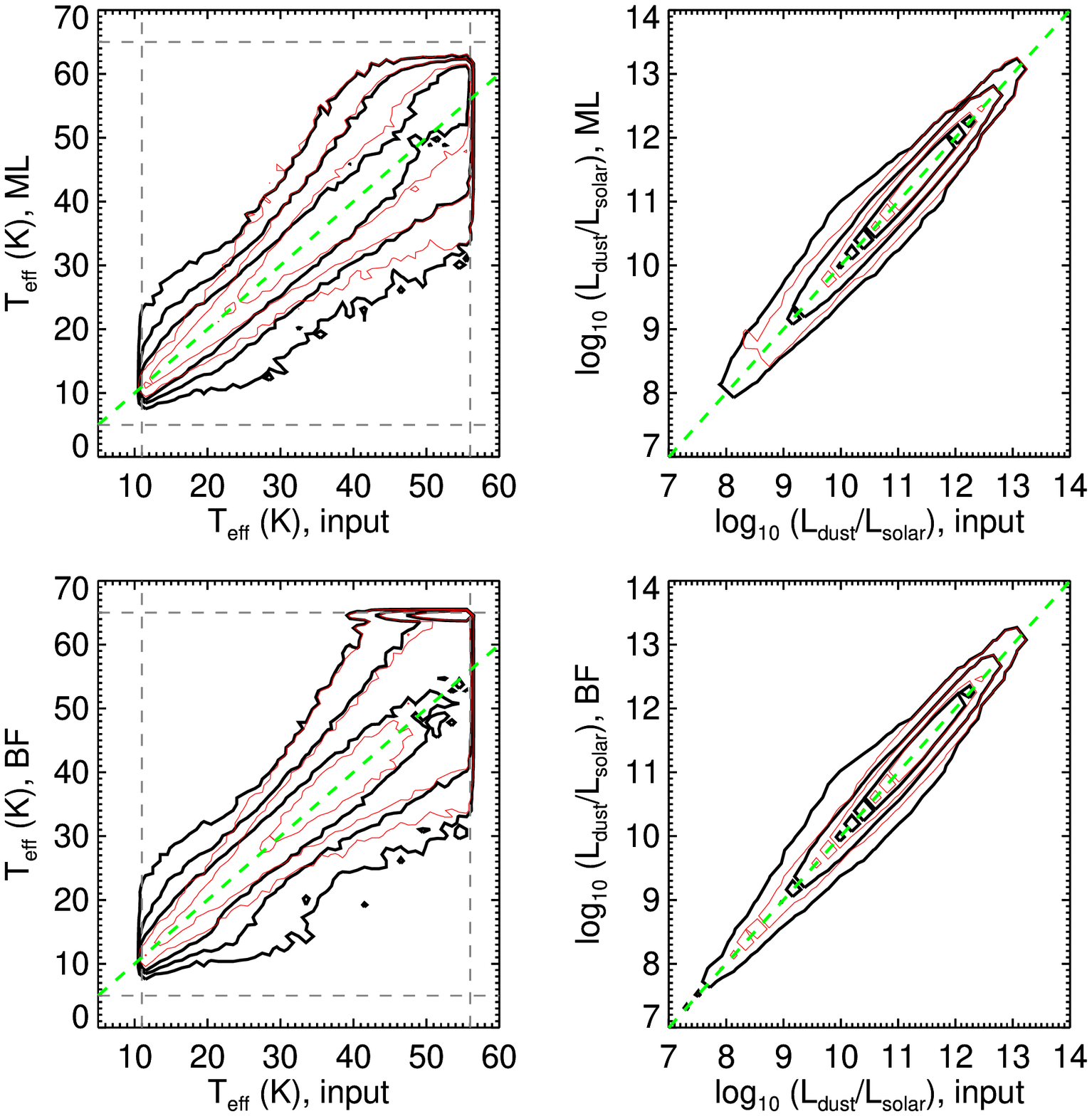}}
  \subfigure[``No PACS'' simulation]{\includegraphics[width=0.33\textwidth]{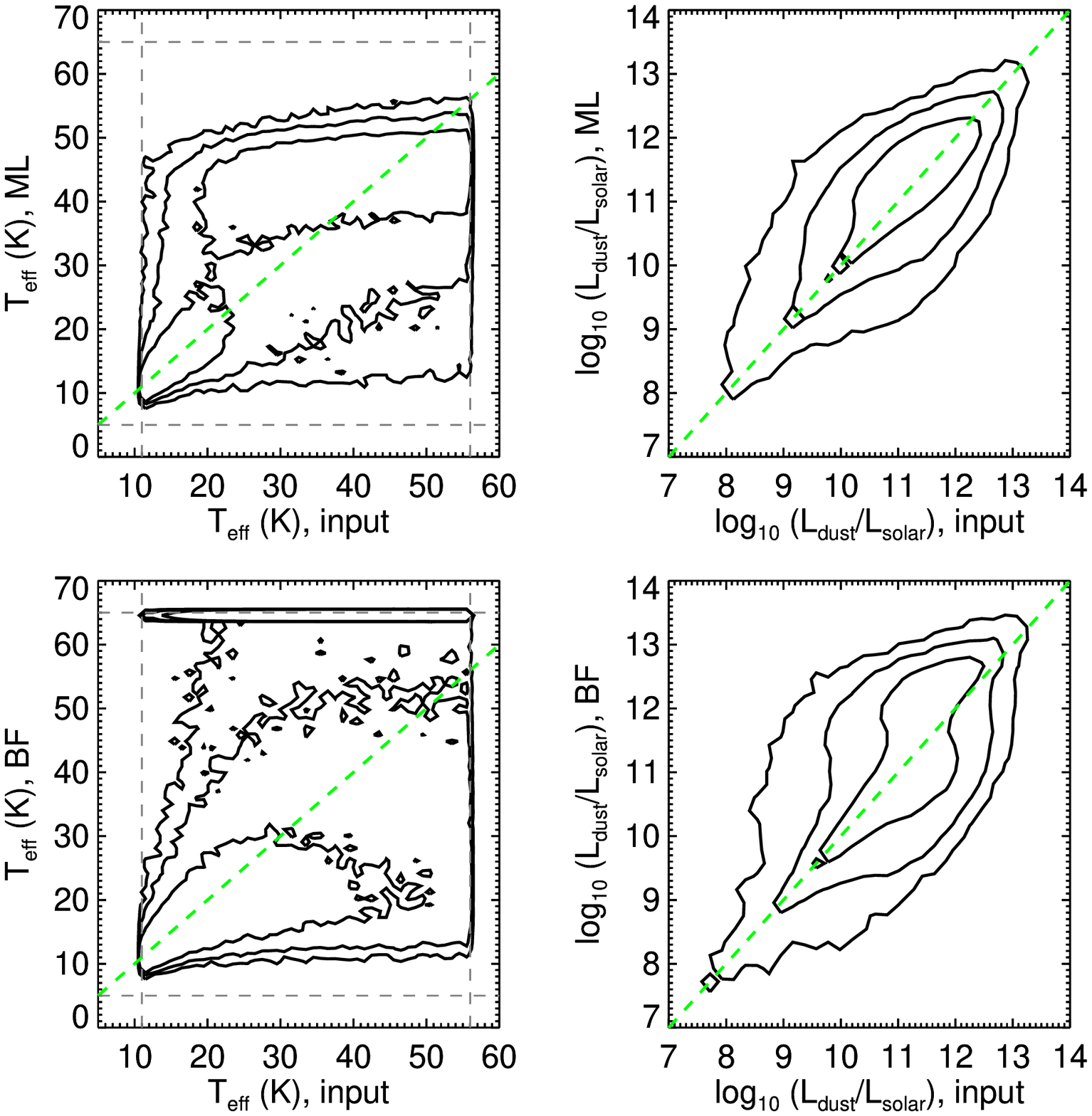}}
  \caption{The results of comparing the input and output
    values for temperature and dust luminosity in our fixed $\beta =
    1.82$, ``Deep PACS'' (a), ``\hatlas-like'' (b) and ``no PACS''
    (c) simulations, from left to right. The top row shows the
    median-likelihood values for each parameter, while the bottom
    row shows the best-fit values, with the thick black contours
    indicating the regions that bound 68.3, 95.5 and 99.7 per cent
    of the full set of simulated sources, and the thin red contours
    showing the same percentiles for the $\ge 5\sigma$ PACS-detected
    fraction of sources. The dashed green lines indicate equal input
    and output values, while the grey dashed lines in the
    temperature plots indicate the bounds of the input and output
    prior distribution; $11 < \tgrey < 56$\,K and $5 < \tgrey <
    65$\,K, respectively. }
  \label{fig:tin_tout_contours}
\end{figure*}

For each set of simulations, we wanted to determine whether we could
accurately recover the known input temperatures (which have been
assigned at random to each of the 200,000 model sources in our
simulation when generating their intrinsic SEDs) and luminosities
(calculated precisely from the noise-free photometry generated in our
simulation) by applying our fixed-$\beta$ SED fitting technique to the
noisy model photometry. In figures \ref{fig:tin_tout_hists} and
\ref{fig:tin_tout_contours} (for which the colour schemes are detailed
in the figure captions), we show tests for possible bias in three of
the four sets of simulations: the ``deep PACS''\footnote{Here we use
  ``deep PACS'' to imply that the PACS data are more sensitive than in
  the fiducial \hatlas-like simulation. This simulation then provides
  a means to study the impact of more sensitive PACS data on the
  derived temperature and dust luminosity estimates. We do not intend
  to imply that this is a simulation of a more traditional deep-field
  survey.}, the \hatlas-like, and the ``no PACS'' simulations are
shown in sub-figures (a), (b) \& (c), respectively. We do not show the
results for the ``no confusion'' simulation, since the results
returned are barely different from the \hatlas-like values.

\begin{figure}
\centering
\includegraphics[height=0.60\columnwidth]{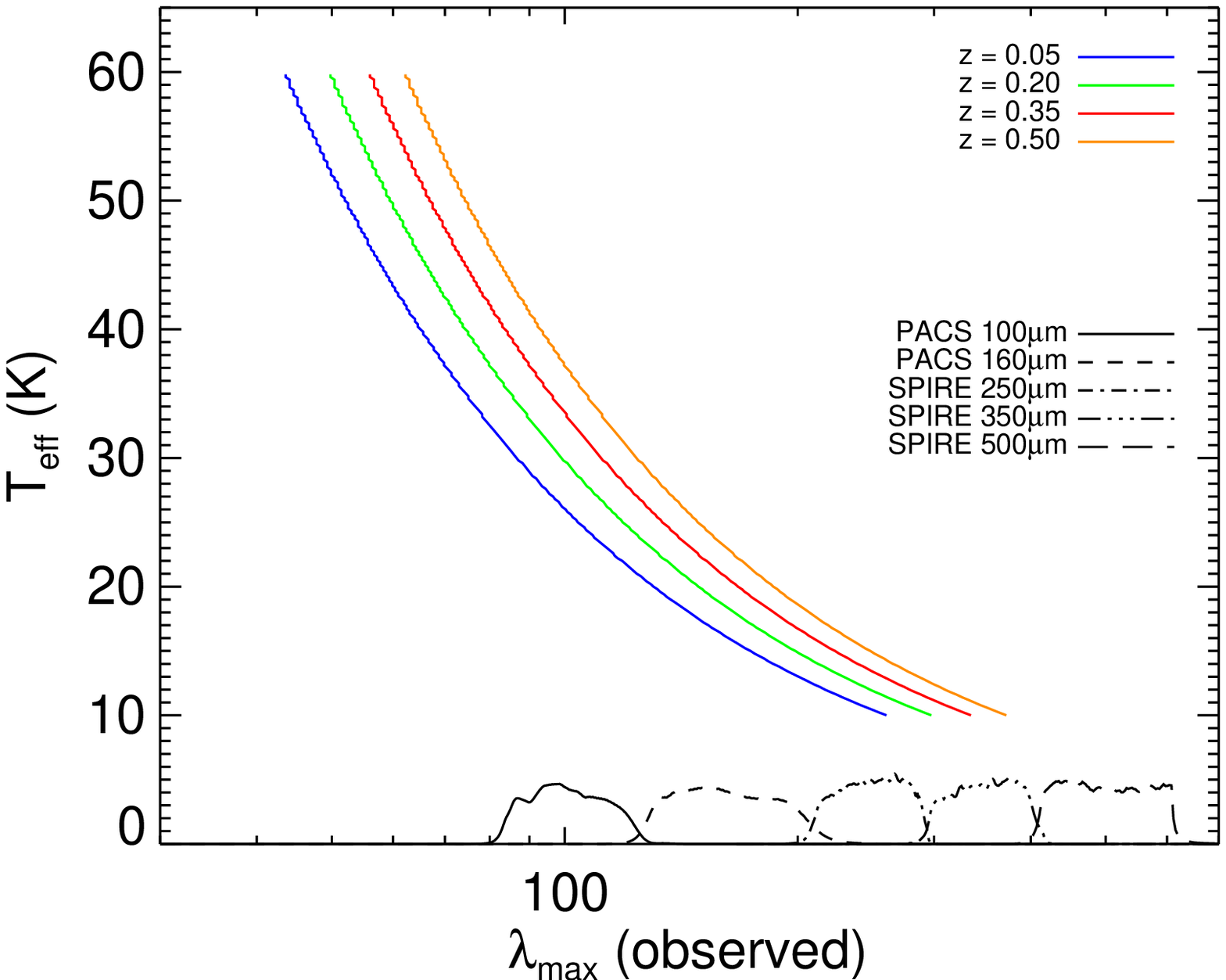}
\caption{The relationship between the wavelength of the peak in the
  dust SED and the modified black-body-equivalent temperature for
  redshifts spanning our sample ($z = 0.05, 0.20, 0.35$ and 0.50),
  assuming $\beta = 1.82$. The PACS (100 and 160\,$\mu$m) and SPIRE
  (250, 350 \&\ 500\,$\mu$m), response curves are overlaid for the
  purposes of comparison.}
\label{fig:lambdamax_t}
\end{figure}

Unsurprisingly, it is clear from comparing the histograms in figure
\ref{fig:tin_tout_hists}\,(a) and (b), as well the corresponding
contour plots in figure \ref{fig:tin_tout_contours}, that more
sensitive PACS data do enable more precise recovery of the input
values of \tgrey\ than we have been able to do in \hatlas, reflected
by the good agreement between the black and green histogram symbols in
figure \ref{fig:tin_tout_hists}, and by the smaller deviations from
the green lines in the \tgrey\ and \ldust\ contour plots (figure
\ref{fig:tin_tout_contours}) using the more sensitive PACS data than
in the \hatlas-like simulation. The systematic differences between the
input and recovered histograms in figure \ref{fig:tin_tout_hists} do
not necessarily imply systematic bias between the input and output
values. The relatively large random errors on the recovered
temperatures for $T > 45$\,K mean that the recovered histogram in
figure \ref{fig:tin_tout_hists}\,(a) smooths over the sharp upper
limit on our flat input temperature distribution. In fact, figure
\ref{fig:tin_tout_contours}\,(a) shows that the recovered temperatures
match the input values very well, despite the peak of the far-IR SED
at these temperatures (the strongest spectral feature in our
broad-band far-infrared photometry) being at wavelengths shorter than
those sampled by the PACS 100\,$\mu$m response curve. To demonstrate
this, we show the variation between the wavelength of the peak in a
model dust SED ($\lambda_{\mathrm{peak}}$) and its isothermal
temperature, along with the observed frame PACS and SPIRE response
functions for a representative range of redshifts and $\beta = 1.82$
in figure \ref{fig:lambdamax_t}.

%\begin{landscape}

%\end{landscape}

Perhaps more surprisingly, in figures \ref{fig:tin_tout_hists}\,(b)
and the left-hand panel of \ref{fig:tin_tout_contours}\,(b) we show
how well we may recover \tgrey\ using the comparatively less sensitive
PACS data from \hatlas\ (though the simulated SPIRE sensitivities are
identical in the two simulations). At $\tgrey\ \ltsim 20$\,K the peak
of the SED falls in the SPIRE 250\,$\mu$m band (at least for the local
galaxies being discussed here) and though there is larger uncertainty
due to the absence of high-significance PACS detections for the
majority of sources -- reflected by the increased spread in the red
and black contours in figure \ref{fig:tin_tout_contours}\,(b) relative
to the ``deep PACS'' plot, figure \ref{fig:tin_tout_contours}\,(a) --
there is little or no bias toward higher or lower temperatures. 

In terms of our ability to recover the input dust luminosity, the
right-hand panels of figure \ref{fig:tin_tout_contours}\,(b) indicate
that the standard deviation of $(\ldust\ -
L_{\mathrm{dust}}^{\mathrm{true}}) \approx 0.19$\,dex across the full
range of \ldust\ is not dissimilar to the value obtained using the
more sensitive PACS simulation shown in figure
\ref{fig:tin_tout_contours}\,(a), $\sim$0.14\,dex.

Whilst for the ``deep-PACS'' and ``\hatlas-like'' simulations there is
little to choose between the median-likelihood and best-fit estimates
of temperature\slash luminosity (solid\slash dashed lines in figure
\ref{fig:tin_tout_hists}, or top and bottom rows in figure
\ref{fig:tin_tout_contours}), in the absence of PACS data this is no
longer the case. Median-likelihood estimates of \tgrey\ are biased
towards values around 45\,K, and the best-fit estimates are similarly
unreliable, frequently biased towards lower values (dashed lines) or
falling on the upper bound of the temperature prior. The flat stacked
temperature PDF for these galaxies at $\tgrey \gtsim 20$\,K, shown as
the purple dot-dot-dashed line in figure
\ref{fig:tin_tout_hists}\,(c), combined with the large peak in the
median-likelihood values around $\tgrey \approx 45$\,K, reflect the
weakness of our temperature constraints in the absence of PACS data.

The severity of the temperature bias in the absence of PACS data is
even more apparent in the figure \ref{fig:tin_tout_contours}\,(c), and
though the ratio of median-likelihood \ldust\ to
$L_{\mathrm{dust}}^{\mathrm{true}}$ is still centred about unity, the
RMS uncertainty is now $\sim$0.5\,dex. The best-fit values for
\ldust\ in the absence of PACS data are frequently biased high at the
highest luminosities, due to the best-fit temperatures falling on the
hot bounds of the prior. It is possible that this situation may
improved upon using more information \citep[e.g. an energy balance
  fitting method; ][]{burgarella05,dacunha08,smith12}, or assuming a
standard temperature for those galaxies with best-fit temperatures
falling at the upper bound of the prior, but the contrast between the
results derived when including and neglecting the PACS data emphasizes
their importance for recovering dust temperatures and luminosities
when fitting a simple modified black-body model, even though sources
may not be formally significantly detected (e.g. $> 3\sigma$). This
result complements the results of \citet{gordon10} and
\citet{skibba11}, who highlighted the importance of SPIRE data for
determining the temperature of the coldest dust; here we stress the
importance of {\em both} sets of data for determining temperatures
across the full range.

\subsection{Temperature sensitivity as a function of redshift}

\begin{figure*}
  \centering 
  \subfigure[``Deep PACS'', median-likelihood]{\includegraphics[width=0.33\textwidth]{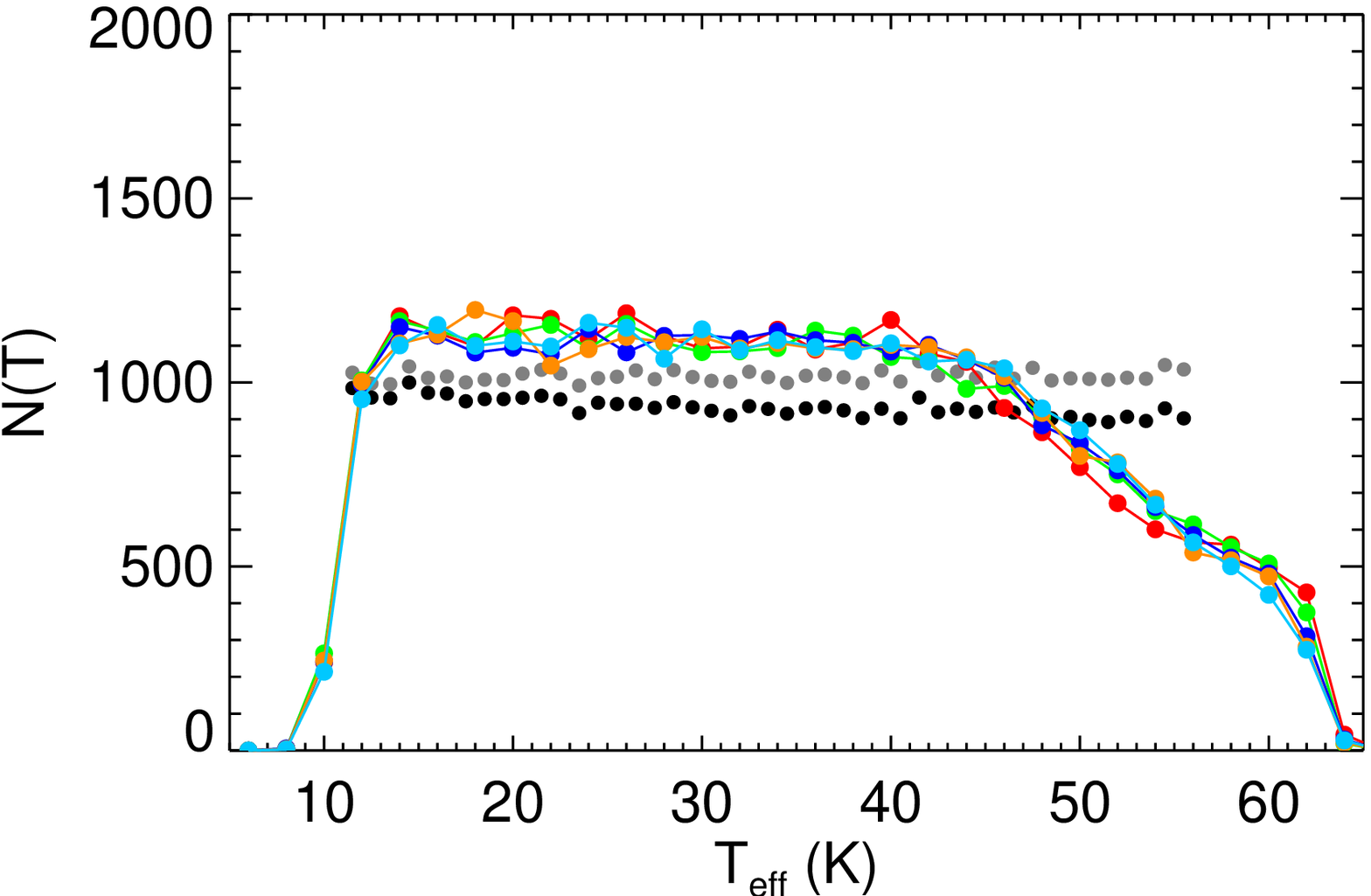}}
  \subfigure[``\hatlas-like'', median-likelihood]{\includegraphics[width=0.33\textwidth]{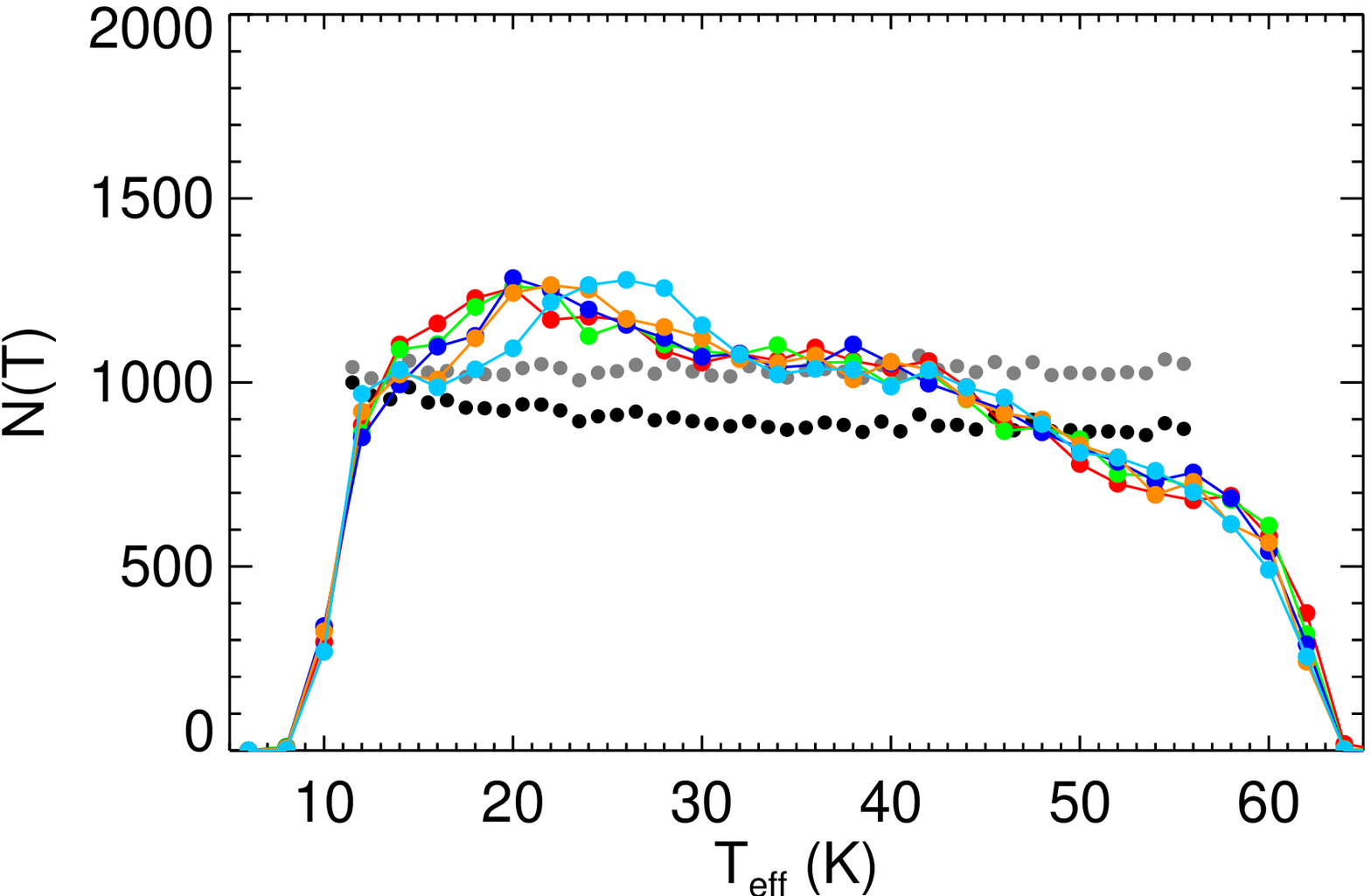}}
  \subfigure[``No PACS'', median-likelihood]{\includegraphics[width=0.33\textwidth]{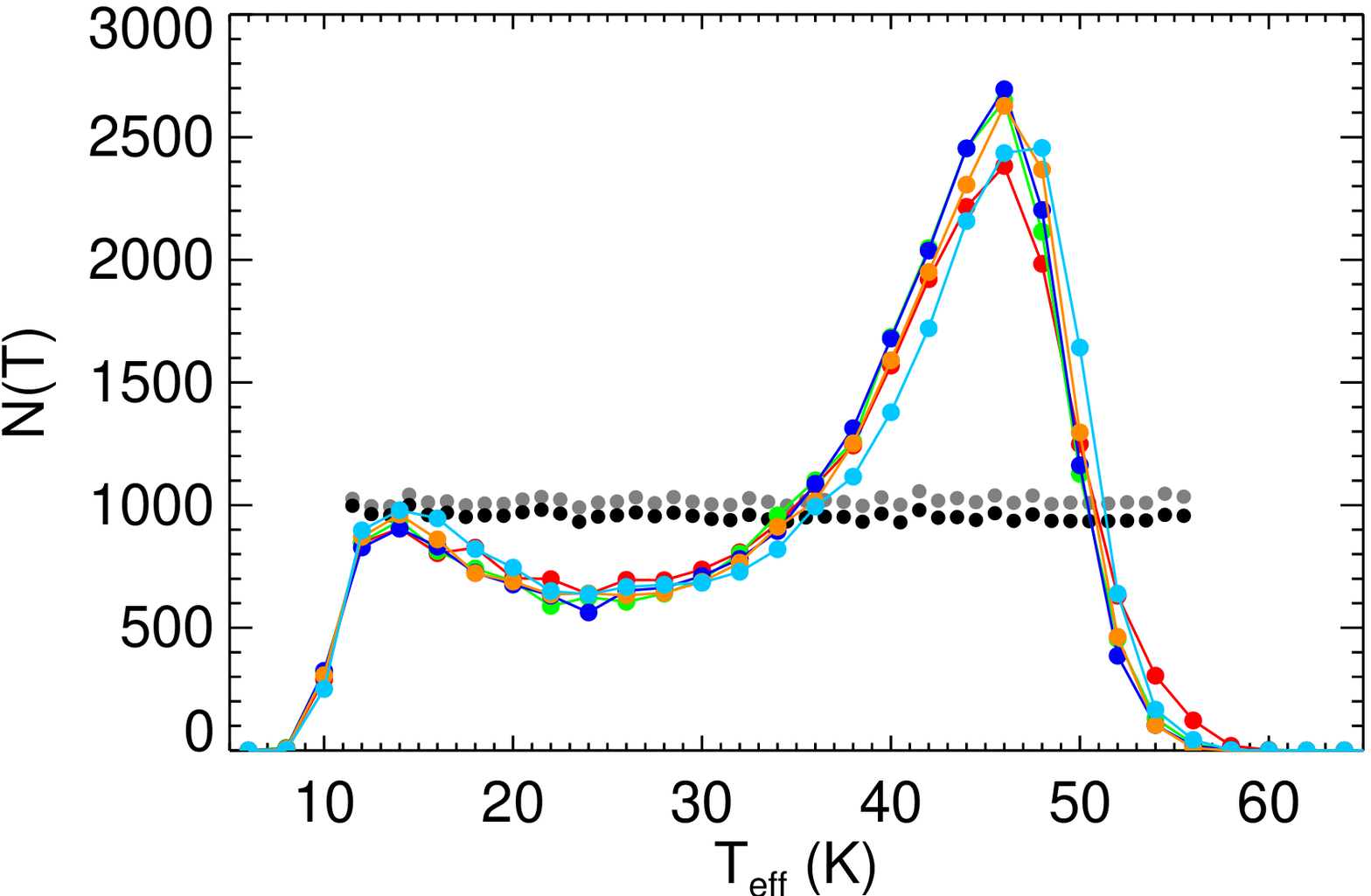}}
  \subfigure[``Deep PACS'', best-fit]{\includegraphics[width=0.33\textwidth]{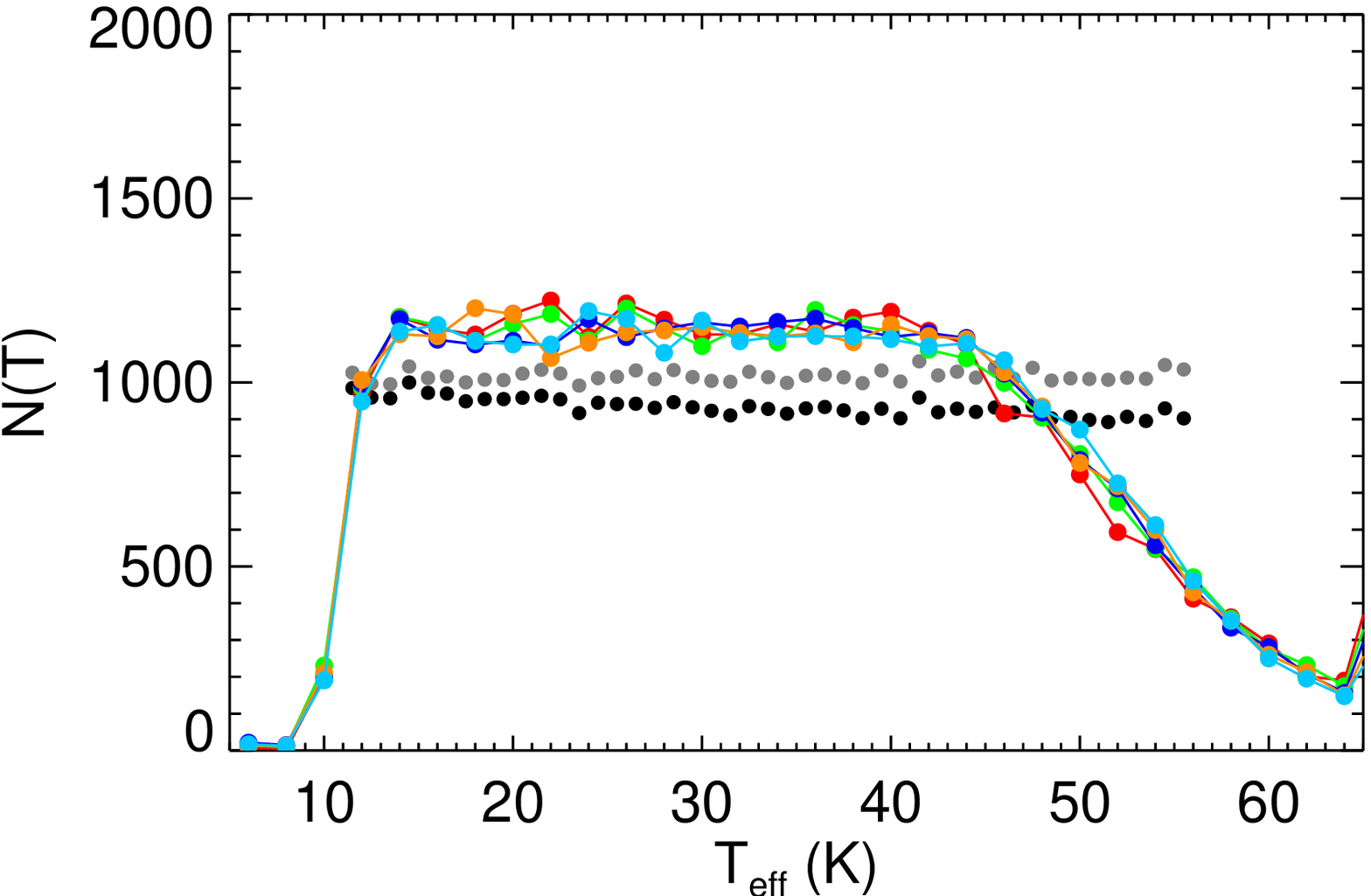}}
  \subfigure[``\hatlas-like'', best-fit]{\includegraphics[width=0.33\textwidth]{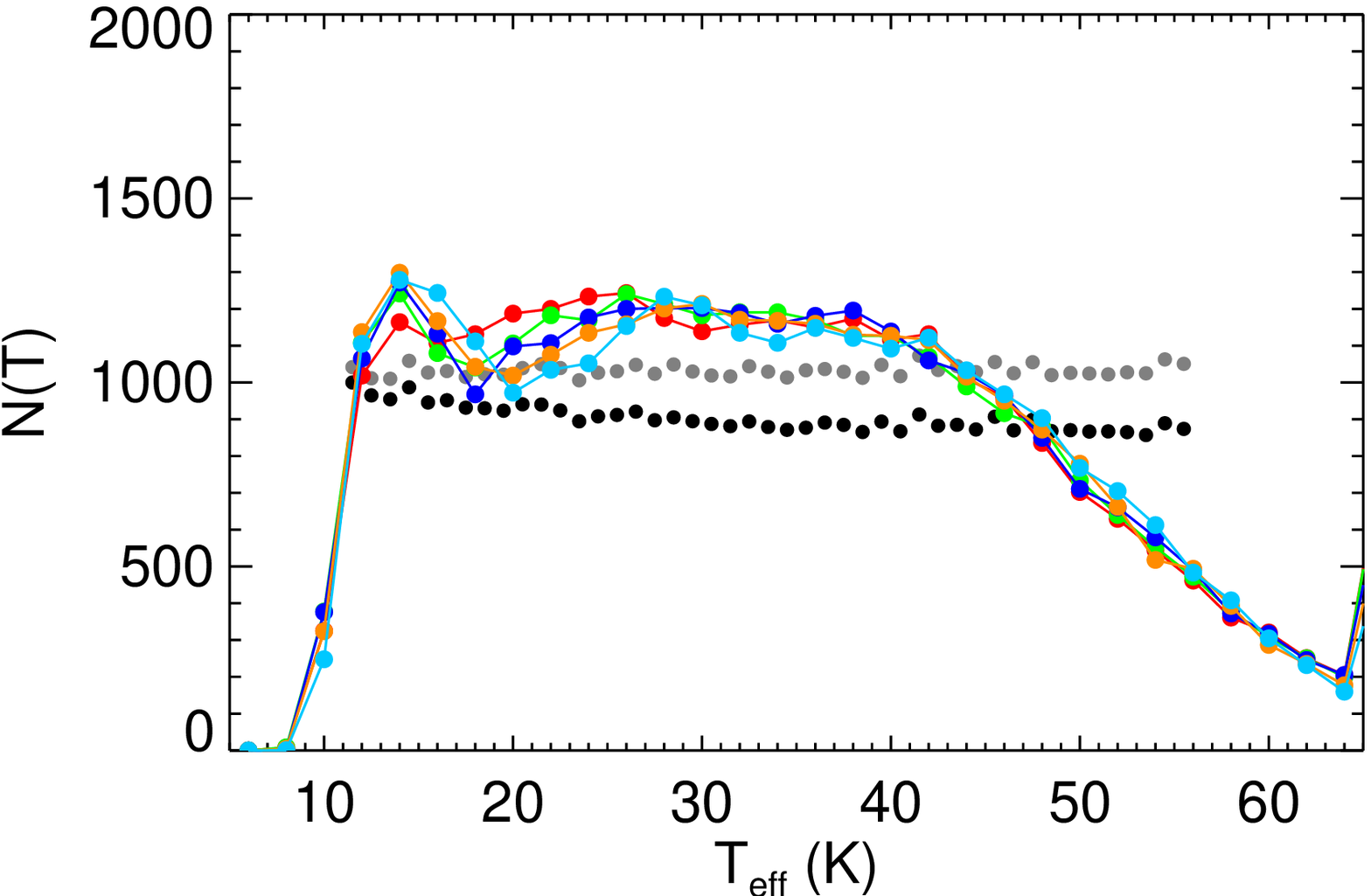}}
  \subfigure[``No PACS'', best-fit]{\includegraphics[width=0.33\textwidth]{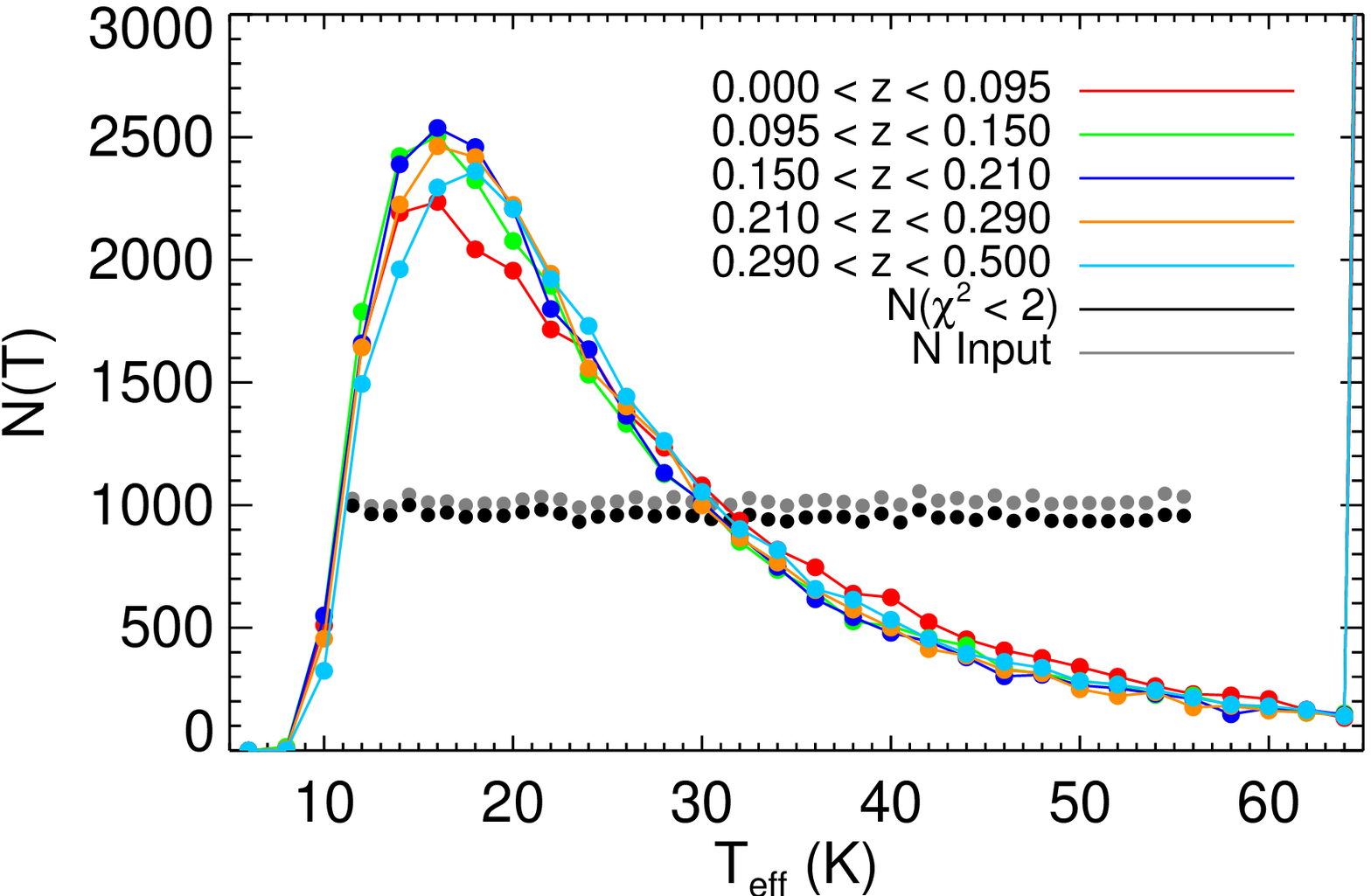}}
  \caption{The variation in the recovered simulated temperatures
    assuming $\beta = 1.82$ in bins of redshift for all sources in our
    simulation (i.e. we make no SNR cuts apart from the 250\,$\mu$m
    selection. The left-hand column shows the results for the ``Deep
    PACS'' simulation, the central column the ``\hatlas-like'', and
    the right-hand column showing the ``no PACS'' simulation
    results. The top row shows the median-likelihood values for each
    simulation, while the bottom row shows the best-fit results. The
    different coloured histograms correspond to the five different
    redshift bins in each simulation, each containing approximately
    equal numbers of sources, with the histograms having been rescaled
    in the vertical direction for the purposes of comparison. The
    redshift ranges for each of the five bins are shown in the legend
    on the bottom-right figure, and are identical in each panel. The
    grey histogram shows the input values, which have been chosen at
    random, while the black histogram shows the input temperatures for
    those sources which are recovered with good fits (i.e. they have a
    reduced $\chi^2 < 2.0$). }
  \label{fig:tsens_redshift}
\end{figure*}

In figure \ref{fig:tsens_redshift} we show histograms of \tgrey,
similar to those in figure \ref{fig:tin_tout_hists}, but in bins of
redshift, such that there are approximately equal numbers of model
galaxies in each bin (the bounds of the bins are $z = 0.095, 0.150,
0.210, 0.290, 0.500$). We show the results for the ``deep PACS'',
``\hatlas-like'' and ``no PACS'' simulations in the left, centre and
right-hand columns, respectively, with the median-likelihood values in
the top row and the best-fit values in the bottom row. As noted
previously, the flat input distribution of temperatures in our
simulation enables us to search for fitting bias. In this case, it
enables us to see whether our temperature sensitivity varies as a
function of redshift due to e.g. the variation in rest-frame
wavelengths being sampled. Our simulation is not intended to model the
\hatlas\ 250\,$\mu$m selection function or the real \hatlas\ coverage
of the range in $\ldust -\tgrey -z$.

Each histogram has been rescaled in the vertical direction for the
purposes of comparison. We overlay the histograms of recovered values
in each redshift bin on the flat input distribution (in grey) and on
the input distribution of galaxies that have good fits when they are
recovered (i.e. reduced $\chi^2 < 2.0$; in black). From comparing the
coloured histograms in each simulation, it is clear that there is
little - if any - evidence for temperature fitting bias that varies as
a function of redshift. Figure \ref{fig:tsens_redshift} also
reinforces the idea that there is little difference between the
best-fit values and the median-likelihood values in the presence of
PACS data. In their absence, the difference is stark, with best-fit
values biased towards cold values ($\tgrey \approx 16$\,K), and
median-likelihood values having a peak near 45\,K, (though the
severity of this bias might be improved with a more
physically-motivated choice of temperature prior).

\section{Fixed emissivity index properties of galaxies in \hatlas}
\label{sec:fixedbeta_results}

\begin{figure*}
\centering
\subfigure{\includegraphics[width=0.90\columnwidth]{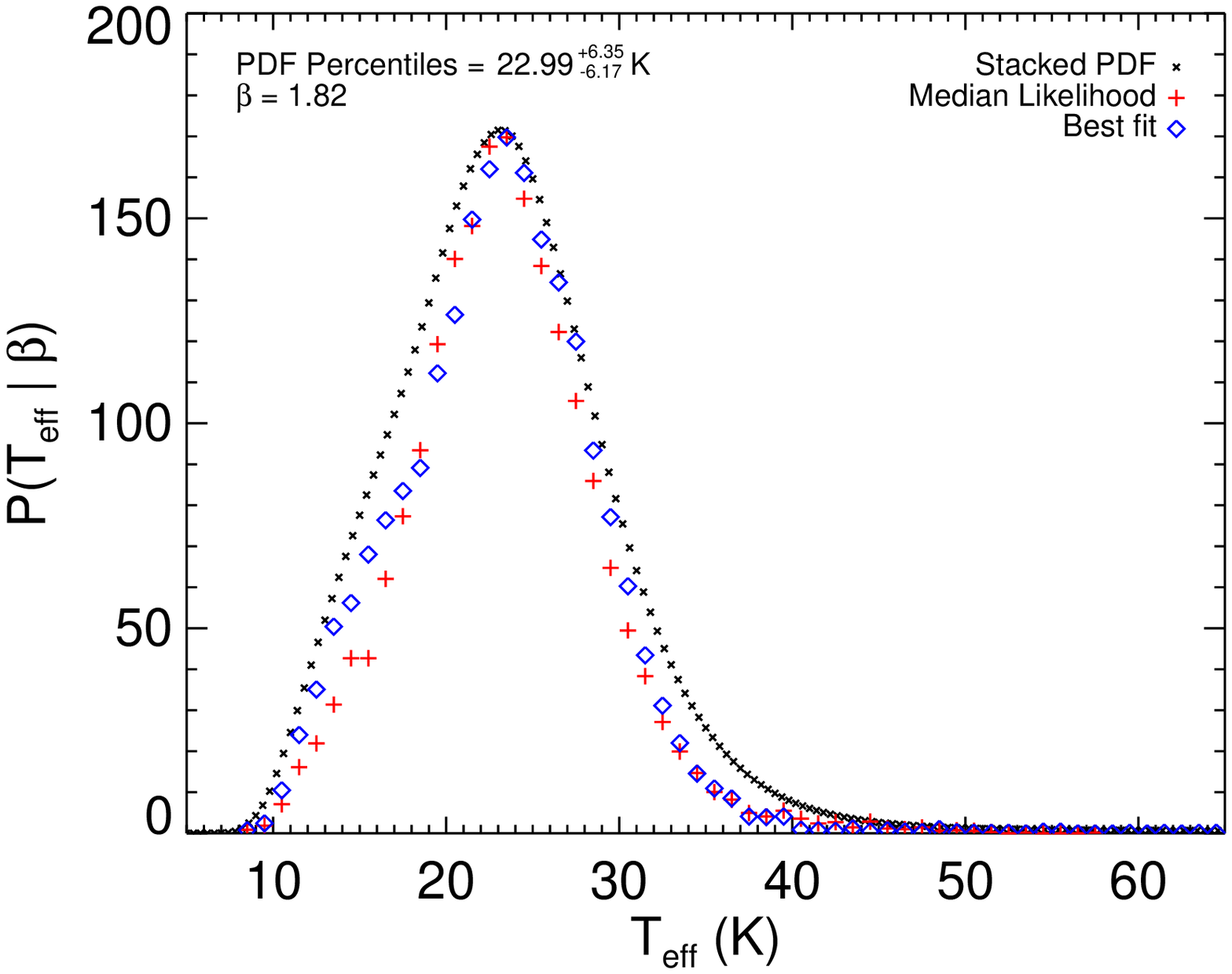}}
\subfigure{\includegraphics[width=0.90\columnwidth]{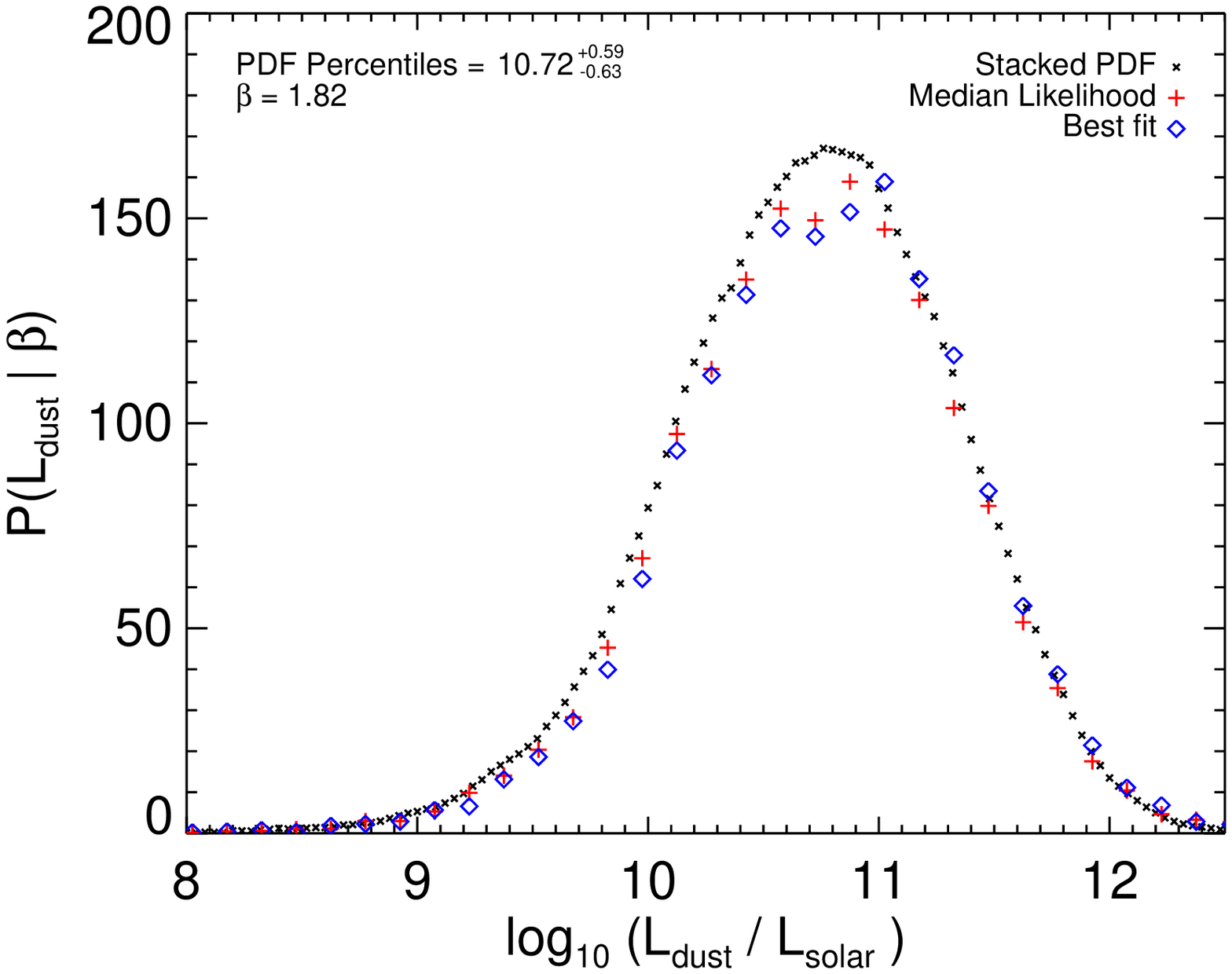}}
\caption{Left: The stacked PDF of $\beta = 1.82$ isothermal
  temperature for 250\,$\mu$m-selected galaxies in \hatlas\ (black
  crosses) with the median-likelihood and best-fit estimates overlaid
  as the red crosses and blue diamonds, respectively. Right: Stacked
  PDF for \ldust\ and histogram of median-likelihood and best-fit
  values in the same colour scheme.}
\label{fig:ldust_tgrey_hatlas}
\end{figure*}

The distributions of \tgrey\ and \ldust\ for the 13,203 of the 13,826
galaxies with $\chi^2$ consistent with the isothermal model are shown
in figure \ref{fig:ldust_tgrey_hatlas}. Clearly, these results --
particularly for the temperature estimates -- show strong dependence
upon our choice of $\beta$. However, even with fixed $\beta$, the
distribution of temperatures in \hatlas\ is broad, with median $23.0
\pm 0.1$\,K and uncertainty between the $16^{\mathrm{th}}$ and
$84^{\mathrm{th}}$ percentiles of $\sim \pm 6.3$\,K according to the
stacked \tgrey\ PDF (black crosses in figure
\ref{fig:ldust_tgrey_hatlas}). We also show the median-likelihood and
best-fit values, as red crosses and blue diamonds, respectively. Our
value of $\tgrey\ = 23.0 \pm 0.1$\,K compares well with the $26.1 \pm
3.5$\,K estimated in \citealt{smith12} for their ``PACS-complete''
sample assuming $\beta = 1.5$, and with other values in the
literature, particularly once the different values of $\beta$ and
sample selections used in these studies are taken in to account
(e.g. \citealt[][who derive $23\pm7$, $21.1 \pm 0.8$, and 20.0\,K
  respectively, assuming $\beta = 2.0$]{dye10,auld12,davies12}).

According to the PDF shown in figure \ref{fig:ldust_tgrey_hatlas}, the
median dust luminosity of a 250\,$\mu$m source in \hatlas\ is
$\log_{10} (\ldust \slash L_\odot) = 10.72 \pm 0.05$ with an
uncertainty of 0.61\,dex.  Using this stacked PDF, we estimate that
while the majority ($\sim 55$ per cent) of \hatlas\ galaxies have far
infrared luminosities in the range classified as star forming
($10^{10} < \log_{10}\left(\ldust \slash L_\odot\right) < 10^{11}$), a
substantial fraction of \hatlas\ galaxies ($\sim 32$ per cent) fall in
the Luminous Infrared Galaxy (LIRG) category using our dust SED
parametrisation. The rest of the population comprises galaxies in the
Normal Infrared Galaxy (NIRG; $\sim$12 per cent of the total) and
Ultra-Luminous Infrared Galaxy (ULIRG; $\sim$1 per cent) categories.

\begin{figure*}
\includegraphics[width=0.90\textwidth]{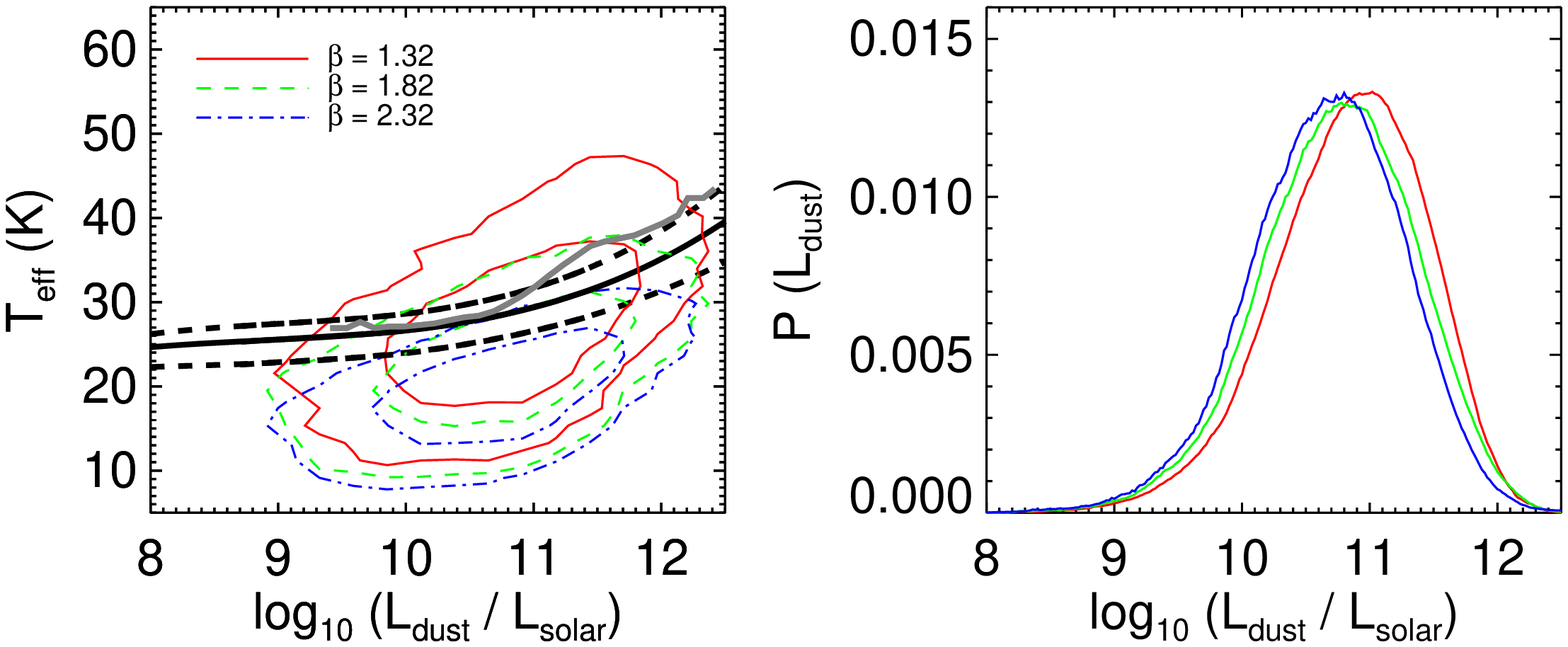}
\caption{{\bf Left:} The effects of choosing different values of
  emissivity index on the derived values of \tgrey\ and \ldust\ using
  our simple fixed-$\beta$ SED fitting code on the \hatlas\ phase 1
  data set. The distribution of the values derived assuming $\beta =
  1.32$, 1.82 and 2.32 are shown as the red, green and blue contours
  respectively, with the best-fit relationships from \citet{chapman03}
  and \citet{hwang10} overlaid in black and grey, respectively. The
  contour levels show the regions enclosed by 68.3 and 95.5 per cent
  of the data at each value of $\beta$. {\bf Right:} stacked PDFs of
  \ldust\ recovered using the same three values and in the same colour
  scheme.}
\label{fig:changing_beta}
\end{figure*}

As we mentioned in section \ref{sec:intro}, it is interesting to probe
the relationship between \ldust\ and \tgrey\ in \hatlas\ galaxies, to
study the implications for deriving luminosity or dust-mass functions,
for example. The emissivity index has a strong influence on this
relationship, as the left hand panel of figure \ref{fig:changing_beta}
shows. Whilst the implications of the choice of $\beta$ for the
derived values of \ldust\ are comparatively modest (using $\beta =
1.32$ or 2.32 -- i.e. $\Delta \beta = 0.5$ -- rather than our fiducial
value of $\beta = 1.82$ changes our estimates of \ldust\ by
approximately 0.05\,dex, as shown in the PDFs in the right panel of
figure \ref{fig:changing_beta}), the impact on the derived values of
\tgrey\ is much larger. This is well demonstrated by the difference
between the red, green and blue contours in the left-hand panel of
figure \ref{fig:changing_beta}, which represent the results of fitting
modified black-body models to the same sample using $\beta = 1.32$,
1.82 \&\ 2.32, respectively\footnote{The median-likelihood estimates
  of \teff\ are $27.1\pm 0.1$, $23.0\pm 0.1$ and $20.0\pm 0.1$\,K for
  the three input values of $\beta$, respectively, while the
  corresponding estimates for $\log_{10} (\ldust \slash L_\odot)$ are
  10.78, 10.72, 10.68, with 0.05\,dex uncertainty on each estimate.}.
 
This range of $\beta$ values is perhaps slightly larger than is
typically suggested in the literature, but nevertheless it is clear
that even at fixed $\beta = 1.82$ (our best estimate, derived in
section \ref{sec:dust_properties}), the 1\,$\sigma$ spread of
\tgrey\ values for a given \ldust\ is likely larger than found by the
previously published ``L-T'' relations in \citet{chapman03},
\citet{hwang10} or \citet{roseboom12}.\footnote{\citet{chapman03} and
  \citet{roseboom12} assume a fixed $\beta = 1.8$ mapping between
  far-IR colour and temperature, while \citet{hwang10} assume $\beta =
  1.5$ in their fitting.} The difference is highlighted by the dashed
green contours in figure \ref{fig:changing_beta}, indicating the
regions that bound 68.3 and 95.5 percent of the \hatlas\ sources, as
compared with the dashed black lines representing the interquartile
range of the values obtained in \citet{chapman03}. Though there is
overlap between our $1\,\sigma$ contours and the interquartile-range
of \citet{chapman03}, our results indicate the presence of a larger
population of cold galaxies detected by \hatlas. As noted by
\citet{rahmati11}, such an increase may be necessary to reproduce the
far-IR source counts in the wavelength regime sampled by SPIRE. Though
dependent upon the choice of $\beta$, we note that an apparent
``L-T''-relation remains irrespective of which value is chosen, albeit
with greater spread in temperature for a given luminosity than has
been previously noted.

\section{Conclusions}
\label{sec:conclusions}

By using the \hatlas\ phase 1 catalogue over $\sim 161$\,deg$^2$ (and
an ancillary suite of simulations based on \hatlas) we show that using
simple $\chi^2$ SED fitting to recover the intrinsic dust properties
of far-infrared sources based on noisy observational data leads to the
introduction of an artificial anti-correlation between isothermal
temperature and emissivity index, confirming the previous results of
\citet{shetty09a} and \citet{kelly12}. The strength of the effect is
such that individual estimates of \tgrey\ and $\beta$ are barely
correlated with their known input values when $\beta$ is allowed to
vary, though we may still derive reasonable estimates of the total
dust luminosity between 8-1000\,$\mu$m, commonly used as a star
formation rate indicator. We have shown that this artificial
anti-correlation persists even when using only those objects with the
highest significance detections in each of our five far-IR bands, as
has been commonly proposed in the literature as a means of mitigating
the impact of this degeneracy.

Since a fixed emissivity-index model is likely to remain the standard
model for studying far-IR galaxy SEDs for the foreseeable future, we
use a simple method to estimate our best estimates of the global mean
temperature and emissivity index for galaxies in \hatlas. We find that
the best values to describe the dust in local ($z < 0.5$) sources are
$\tgrey\ = 23.5 \pm 0.1$\,K and $\beta = 1.82 \pm 0.02$, where we
correct for residual bias in our fitting by making the assumptions
that temperature and emissivity index have no intrinsic correlation,
and that the intrinsic distribution of values is Gaussian. 

By splitting our sample into bins containing approximately equal
numbers of sources based on their redshifts, we recover increasing
mean temperature for \hatlas\ galaxies as a function of redshift
\citep[as expected from previous studies which have found that
  luminosity and temperature are correlated, though sample selection
  has also played a role in this correlation; see e.g.][for a
  comprehensive discussion of selection effects on 60 and
  250\,$\mu$m-selected surveys]{symeonidis13}. Furthermore, we find
tentative evidence, at the 2.8\,$\sigma$ level, that the population
mean value of the emissivity index positively evolves with redshift,
assuming a linear relationship between the two parameters. We leave an
investigation of the individual galaxy properties for a future study,
since the artificial anti-correlation discussed in section
\ref{sec:method} \citep[and noted by other authors,
  e.g.][]{shetty09a,kelly12,veneziani13} precludes such analysis using
these techniques.

It is difficult to reconcile our best-fit results with those of
\citet{kelly12}, who suggest emissivity indices $\beta > 2$ for
temperatures between $12<T<15$\,K, and weak positive correlation
between the two (derived using a Spearman's rank method), though such
high values for $\beta$ have been observed by other studies of
galactic sources \citep[e.g.][]{hill06,miyake93,kuan96} and attributed
to grain growth in the central region, or the presence of ice-coated
dust grains. Some simple explanations for this apparent discrepancy
might be that the dust properties of the Bok globule CB244 may not be
representative of the extragalactic population, that any correlation
between temperature and emissivity index may be a more complicated
function of the dust temperature, that dust emissivity may not simply
vary as a power law function of frequency, or some combination of
these. Since it is all but certain that the effects of superposing
different dust clouds upon a line of sight affect all of these
observations \citep[e.g.][]{shetty09a,veneziani13}, in this paper we
refer to effective temperatures and emissivity indices, which describe
the emergent integrated spectrum, and are useful for our purposes. It
is likely to be extremely complicated to infer the intrinsic
properties of the individual dust populations within these sources
using these data.

Based on our best estimate of \beff\ in \hatlas, and on a further
suite of ancillary simulations with varying sensitivity in the PACS
bands centred on 100 and 160\,$\mu$m, we find that our ability to
derive fixed-$\beta$ estimates of isothermal temperature and dust
luminosity using simple $\chi^2$ fitting is dramatically improved by
including the \hatlas\ PACS data in our analysis, even though these
data are considerably less sensitive than the \hatlas\ SPIRE data, and
may not be formally significant detections (e.g. $> 3\,\sigma$). We
also show that our ability to determine fixed $\beta$ dust
temperatures in \hatlas\ shows only weak dependence on the redshift of
the galaxy in question, at least out to $z < 0.5$.

Finally, we determine that the median $\beta = 1.82$ dust luminosity
of 250\,$\mu$m selected galaxies in \hatlas\ at $z < 0.5$ is
$\log_{10} (\ldust \slash L_\odot) = 10.72 \pm 0.05$, though the
choice of $\beta$ has little influence on this value, including
whether it is allowed to vary or is held fixed. We find that while the
majority of \hatlas\ galaxies ($\sim 54$\,per cent) fall in the
star-forming category, a substantial minority ($\sim 31$ per cent) are
classified as LIRGs according to their dust luminosity.

\appendix

\section*{Acknowledgments}
{The {\it Herschel}-ATLAS is a project with {\it Herschel}, which is
  an ESA space observatory with science instruments provided by
  European-led Principal Investigator consortia and with important
  participation from NASA. The {\it H}-ATLAS website is
  \url{http://www.h-atlas.org/}. GAMA is a joint European-Australasian
  project based around a spectroscopic campaign using the
  Anglo-Australian Telescope. The GAMA input catalogue is based on
  data taken from the Sloan Digital Sky Survey and the UKIRT Infrared
  Deep Sky Survey. Complementary imaging of the GAMA regions is being
  obtained by a number of independent survey programs including {\it
    GALEX} MIS, VST KIDS, VISTA VIKING, WISE, GMRT and ASKAP providing
  UV to radio coverage. GAMA is funded by the STFC (UK), the ARC
  (Australia), the AAO, and the participating institutions. The GAMA
  website is \url{http://www.gama-survey.org/}. This work used data
  from the UKIDSS DR5 and the SDSS DR7. The UKIDSS project is defined
  in \citet{lawrence07} and uses the UKIRT Wide Field Camera
  \citep[WFCAM][]{casali07}. Funding for the SDSS and SDSS-II has been
  provided by the Alfred P. Sloan Foundation, the Participating
  Institutions, The National Science Foundation, the U.S. Department
  of Energy, the National Aeronautics and Space Administration, the
  Japanese Monbukagakusho, the Max Planck Society and the Higher
  Education Funding Council for England. MJJ is partly supported by
  the South African Square Kilometer Array Project. The Italian group
  acknowledges partial financial support from ASI\slash INAF agreement
  n. I/009/10/0. J.G.N. acknowledges financial support from Spanish
  CSIC for a JAE-DOC fellowship and partial financial support from the
  Spanish Ministerio de Ciencia e Innovacion project
  AYA2010-21766-C03-01.}

\bibliographystyle{mn2e}\bibliography{lt_paperrefs}

\bsp

\label{lastpage}

\end{document}